\documentclass[aps,amsmath,prx,a4paper,floatfix,notitlepage,twocolumn, superscriptaddress, longbibliography]{revtex4-1}
\usepackage{graphicx}
\usepackage{amsmath}
\usepackage{amsfonts}
\usepackage{amssymb}
\usepackage{bm}
\usepackage{color}
\usepackage[linktocpage=true]{hyperref}
\newcommand{\edit}[1]{{#1}}

\begin{document} 
\title{Inverse Funnel Effect of Excitons in Strained Black Phosphorus}
\author{Pablo San-Jose}
\affiliation{Instituto de Ciencia de Materiales de Madrid, ICMM-CSIC, Cantoblanco, E-28049 Madrid, Spain}
\author{Vincenzo Parente}
\affiliation{Instituto Madrile\~no de Estudios Avanzados en Nanociencia (IMDEA-nanociencia), Cantoblanco, E-28049 Madrid, Spain}
\affiliation{Instituto de Ciencia de Materiales de Madrid, ICMM-CSIC, Cantoblanco, E-28049 Madrid, Spain}
\author{Francisco Guinea}
\affiliation{School of Physics and Astronomy, University of Manchester, Oxford Road, Manchester M13 9PL, UK}
\affiliation{Instituto Madrile\~no de Estudios Avanzados en Nanociencia (IMDEA-nanociencia), Cantoblanco, E-28049 Madrid, Spain}
\author{Rafael Rold\'an}
\affiliation{Instituto de Ciencia de Materiales de Madrid, ICMM-CSIC, Cantoblanco, E-28049 Madrid, Spain}
\author{Elsa Prada}
\affiliation{Departamento de F\'isica de la Materia Condensada, Condensed Matter Physics Center (IFIMAC), and Instituto Nicol\'as Cabrera, Universidad Aut\'onoma de Madrid, E-28049 Madrid, Spain}

\begin{abstract}
We study the effects of strain on the properties and dynamics of Wannier excitons in monolayer (phosphorene) and few-layer black phosphorus (BP), a promising two-dimensional material for optoelectronic applications due to its high mobility, mechanical strength and strain-tuneable direct band gap. We compare the results to the case of molybdenum disulphide (MoS$_2$) monolayers. We find that the so-called funnel effect, i.e. the possibility of controlling exciton motion by means of inhomogeneous strains, is much stronger in few-layer BP than in MoS$_2$ monolayers and, crucially, is of opposite sign. Instead of excitons accumulating isotropically around regions of high tensile strain like in MoS$_2$, excitons in BP are pushed away from said regions. This \emph{inverse} funnel effect is moreover highly anisotropic, with much larger funnel distances along the armchair crystallographic direction, leading to a directional focusing of exciton flow. A strong inverse funnel effect could enable simpler designs of funnel solar cells, and offer new possibilities for the manipulation and harvesting of light.
\end{abstract}
\maketitle

\section{Introduction}

Two-dimensional (2D) crystals, such as graphene, transition metal dichalcogenides (TMDs) and, more recently, few-layer black phosphorus (BP), have revealed great technological potential thanks in particular to their unique combination of mechanical and optoelectronic properties. On the one hand, many of these atomically thin membranes can withstand unprecedented strains of up to $10 - 25\%$ without plastically deforming or rupturing \cite{Feng:NP12}. This is in stark contrast to most bulk semiconductors that fail mechanically at strains of about $0.1 - 0.4\%$. On the other hand, these materials cover a wide range of optically active (i.e. direct) band gaps. A particularly important 2D crystal in this regard is few-layer BP \cite{Li:NN2014,Koenig:APL2014,Liu:ACS2014,Castellanos-Gomez:2M14}, as it is the only member of the family with a direct gap that covers the range between 0.3 eV and 1.8 eV as the number of layers is decreased. This is a crucial range of energies for many semiconductor technologies \cite{Xia:NC14,Castellanos-Gomez:JPCL15}, including infrared photodetectors \cite{Guo:16}, telecommunications \cite{Youngblood:NP15}, and even photovoltaics \cite{Xia:NP14,Ganesan:APL16,Castellanos-Gomez:NP16}, which could furthermore benefit from BP's high mobilities \cite{Gillgren:2M15,Chen:NC15,Li:NN15}. Finally, several of these 2D crystals exhibit an extraordinarily strong coupling between these two aspects, strain and optical activity \cite{Roldan:JPCM15}. The gap of TMDs monolayers, for example, \emph{decreases} by up to $\sim 1.5\%$ under $1\%$ of uniaxial tension \cite{Chang:PRB13}. Once more, few-layer BP is remarkable in this regard, as it shows one of the strongest modulation of its band gap, ranging from a $\sim 6\%$ (monolayer) \cite{Peng:PRB14} to a $\sim23\%$ (bulk)  \emph{increase} under 1\% of uniaxial tension \cite{Quereda:NL16}. Not only is the strain sensitivity of the BP gap stronger and of opposite sign to that of TMDs, but it is also expected to be anisotropic with the direction of applied uniaxial strain \cite{Fei:NL14}. This bestows few-layer BP with rather unique opportunities in the field of optoelectronics that are only recently beginning to be explored.

\begin{figure}[t]
\includegraphics[width=\columnwidth]{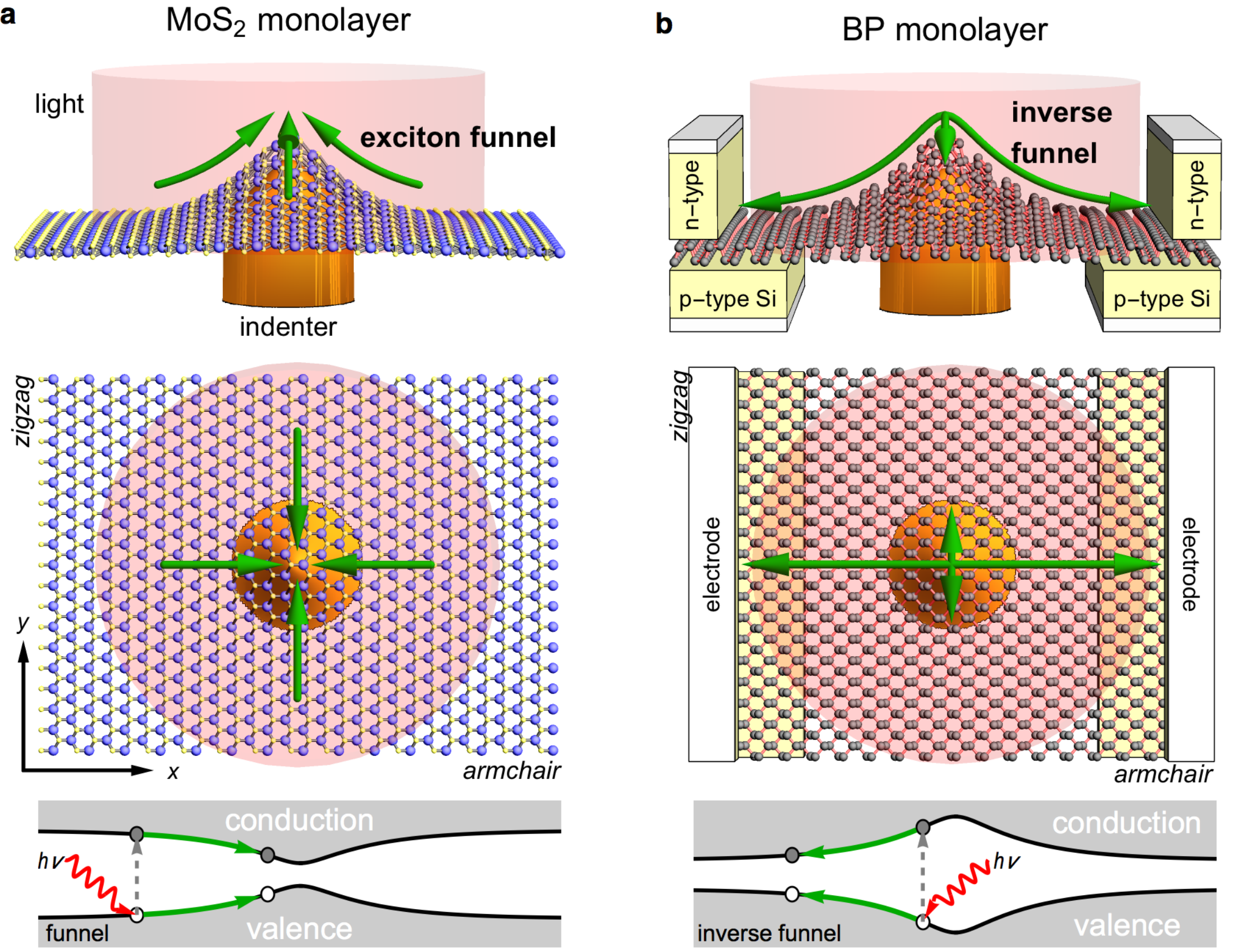}
\caption{Color online: An exciton funnel for MoS$_2$ is represented in (a), where an indenter creates an inhomogeneous strain profile that modulates the gap (bottom), and pushes photogenerated excitons (in green) isotropically towards the center of indentation. In BP (b), the same strain profile creates a stronger, highly anisotropic inverse funnel effect that pushes excitons away from the indentation along the armchair direction. }
\label{fig:funnel}
\end{figure}

A fundamental optoelectronic phenomenon in semiconductors is the generation and recombination of excitons, i.e. particle-hole pairs that become bound by Coulomb interaction and form a state with energy $E_\mathrm{ex}$ inside the semiconductor gap $E_\mathrm{gap}$.
When illuminated by a light source of frequency $\omega>E_\mathrm{gap}/\hbar$, electrons are excited from the valence to the conduction band. The electron and the hole lose energy through several mechanisms \cite{Moody:16}, eventually relaxing to the exciton state in the gap. After a finite lifetime $\tau$, typically longer than the preceeding relaxation, the exciton recombines through the emission of a photon of energy $E_\mathrm{ex}<E_\mathrm{gap}$, producing photoluminescence. A number of 2D crystals have remarkably strong excitonic photoluminescence \cite{Splendiani:NL10,Mak:NN12,Ugeda:NM14,Wang:NN15,Yang:LSA15}, with binding energies $E_b=E_\mathrm{gap}-E_\mathrm{ex}$ typically exceeding those of semiconductors. 

It has been proposed that strain-engineering of the gaps of 2D crystals could be used to efficiently manipulate excitons. By creating a strain gradient, e.g. by localized elastic indentation of the crystal, the exciton energy $E_\mathrm{ex}$ is expected to  vary spatially in a similar way as the gap itself. Feng \textit{et al.} \cite{Feng:NP12} predicted that strain gradients in MoS$_2$ monolayers create a force on (neutral) excitons that pushes them towards the regions of maximum tension (least gap) \footnote{\edit{Note that in gapless graphene, excited electrons and holes give rise to a photocurrent through the thermoelectric effect \cite{Gabor648}. In contrast, the funnelling of (neutral) excitons considered here is driven by the potential gradient associated to the changing gap, and it does not involve charge currents.}}, in what was dubbed an ``exciton funnel'', see Fig. \ref{fig:funnel}a. They argued that in a photovoltaic solar cell, the modulation of the optical gap through strain and the efficient funnelling of excitons to specific locations could lead to significant performance gains when compared to the standard photocarrier diffusion in conventional, fixed-gap cells, even beating the Shockley-Quessier limit \cite{Shockley:JAP61}. Various aspects of exciton funnelling in MoS$_2$ have been explored experimentally \cite{Castellanos-Gomez:NL13, Li:NC15}. More generally, a range of promising applications for strain-engineered optical properties in 2D crystals have been proposed \cite{Jariwala:AN14,Yu:MB14,Li:MB14}. 
 
In this work we study the properties of excitons in strained few-layer BP and their dynamics under strain gradients, and compare them to the case of MoS$_2$ monolayers. We find that BP exhibits a strongly anisotropic \emph{inverse} funnel effect, whereby excitons are efficiently driven away from regions with tensile strains in a specific direction relative to the crystal axis. This behaviour is rare amongst known 2D crystals, and it could prove preferable to the original funnel effect of TMDs, by separating the source of strain and the location of exciton accumulation. An example is the inverse funnel solar cell of Fig. \ref{fig:funnel}b, wherein the optically active regions are strained, and separated from the unstrained regions under the electrodes. We furthermore show that the absolute funnelling efficiency in few-layer BP is potentially far better than in MoS$_2$ monolayers, particularly as the number of layers increases.

\begin{figure}[t]
\includegraphics[width=\columnwidth]{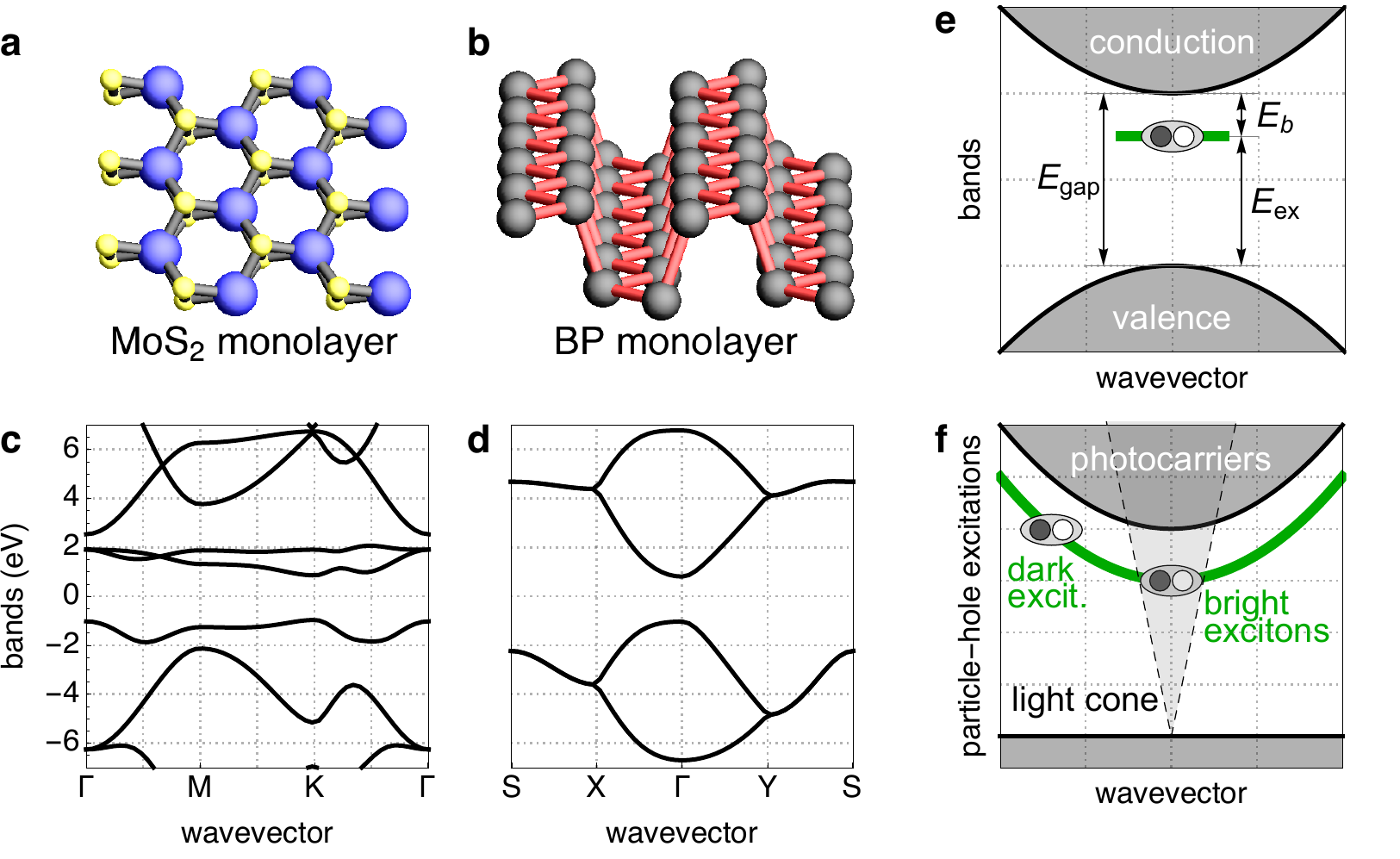}
\caption{Color online: Crystalline structure of MoS$_2$ (a) and BP (b) monolayers, with their corresponding band structures within our tight-binding approach (c,d). (e) Schematic representation of the single-particle energy bands, with an exciton state at energy $E_{ex}$ marked in green. (f) Schematic representation of the particle-hole excitation spectrum. Excitons within the light cone $E<\hbar c|k|$ can decay radiatively, and are thus `bright'.}
\label{fig:bands}
\end{figure}

\section{Formalism} 
The starting point to analyse exciton formation in 2D crystals is an accurate description of the non-interacting, strain-free bandstructure around the gap. To this end we employ a tight-binding description (see Appendix \ref{ap:TB}) carefully fitted to ab-initio calculations, both for few-layer BP (Ref. \cite{Rudenko:PRB15}) and MoS$_2$ (Ref. \cite{Cappelluti:PRB13}). These tight-binding models include $p_z$ phosphorus orbitals for BP, and $p_{x,y,z}$ sulfur orbitals plus the five $d$ molybdenum orbitals for MoS$_2$, see lattices in Figs. \ref{fig:bands}(a,b). The models can be extended to incorporate arbitrary strain profiles, an important advantage over ab-initio approaches. The resulting bands (Fig.  \ref{fig:bands}c,d), gap values and carrier effective masses are consistent with experimental and theoretical results available in the literature. We ignore spin-orbit coupling, which is responsible in MoS$_2$ for a range of interesting spin-dependent phenomena, that are however not essential for the present work.

As expected, in the case of few-layer BP the gap from the tight-binding model is direct and sits at the $\Gamma$ point. Its value ranges from $E_\mathrm{gap}=1.84$ eV for the monolayer (also known as `phosphorene', Fig. \ref{fig:bands}d) to $E_\mathrm{gap}=0.41$ eV for bulk BP \cite{Castellanos-Gomez:JPCL15}.  The MoS$_2$ monolayer gap in our model is around $E_\mathrm{gap}=1.82$ eV \footnote{The gap for MoS$_2$ is underestimated by about $20-25\%$ in our model as compared to most experiments. This is expected, as the tight-binding for MoS$_2$, taken from Ref. \cite{Cappelluti:PRB13}, was fitted to LDA calculations, which are known to underestimate gaps \cite{Perdew:IJQC85,Qiu:PRL13}, as opposed to the GW-LDA used for BP \cite{Rudenko:PRB15}. This in turn is expected to lead to an overestimation of exciton lifetimes and funnel drift lengths in MoS$_2$.}, but sits at the $K$ point (Fig. \ref{fig:bands}c), and becomes indirect for multilayer MoS$_2$ samples. As an indirect gap is much less active optically, many optoelectronic applications of MoS$_2$ are mostly restricted to the monolayer. Another important difference between the bandstructure of the two materials is BP's strong anisotropy of carrier effective masses. While the effective mass of MoS$_2$ for the conduction and valence band is isotropic due to lattice symmetry,  $m^{c,v}_x=m^{c,v}_y\approx 0.5 m_e$, in monolayer BP we have a high anisotropy both in the conduction ($m^c_x\approx 0.2 m_e$, $m^c_y\approx 1.1 m_e$) and in the valence bands ($m^v_x\approx 0.2 m_e$, $m^v_y\approx 3.9 m_e$). (Throughout this work, $x$ refers to the armchair orientation, and $y$ to zigzag, see Fig. \ref{fig:funnel}.) Significant anisotropies persist as the number of layers increases,  a consequence of BP's puckered lattice structure (Fig. \ref{fig:bands}b).  It is crucial to take into account the mass anisotropy when discussing exciton formation, as effective masses directly control their binding energies and spatial dimensions.

The problem of anisotropic excitons has only recently been analysed \cite{Rodin:PRB14,Tran:PRB14,Prada:PRB15,Wang:NN15,Chaves:PRB15}. In the  limit of low exciton density, it is possible to treat the Coulomb interaction between a single electron and a single hole as a two-body problem \cite{Yu:05}. Here we use the theory of Ref. \cite{Prada:PRB15}, where approximate analytical solutions, valid for anisotropic electron and hole masses, were derived. The wave function of an exciton of total momentum $\hbar\vec{Q}$ can be expressed as $\Psi_{ex}(\vec R, \vec r)=e^{i\vec{Q}\cdot\vec{R}}\phi(\vec{r})$, where $\vec R$ is the center-of-mass coordinate, and $\vec r$ is the relative coordinate.
In the effective mass approximation,
the total exciton energy disperses with wavevector $\vec Q$ as $E(\vec Q)=E_\mathrm{ex}+\sum_i\hbar^2Q_i^2/2M_i$, see Fig. \ref{fig:bands}f,  where the total masses are $M_{x,y}=m^{c}_{x,y}+m^{v}_{x,y}$. The function $\phi(\vec{r})$ satisfies the Schr\"odinger equation, with reduced masses $\mu_{x,y}^{-1}=1/m^{c}_{x,y}+1/m^{v}_{x,y}$ and Coulomb interaction $V(\vec{r})$ between the electron and the hole. Its solution yields the binding energy $E_b$ from which $E_\mathrm{ex}=E_\mathrm{gap}-E_b$ (see Fig. \ref{fig:bands}e). As shown in Ref. \cite{Keldysh:JL79}, the Coulomb interaction $V(\vec r)$ in a thin slab of thickness $d$ and with dielectric constant $\varepsilon$, embedded between two dielectric media with constants $\varepsilon_1$ and $\varepsilon_2$, depends on the screening length $r_0=d\varepsilon/(\varepsilon_1+\varepsilon_2)$, which marks the crossover between a logarithmic divergence for $r<r_0$ and the usual $1/r$ behavior for $r>r_0$. For anisotropic materials, one can approximate $\varepsilon=(\varepsilon_x\varepsilon_y\varepsilon_z)^{1/3}$ \cite{Landau:84}. In a suspended BP monolayer ($d=5.24$~\AA, $\varepsilon=10.3$), the screening length is around $r_0\approx 25$ \AA. Since the estimated excitonic radii $a_{x,y}$ in BP monolayer are smaller than $r_0$ (see Table \ref{tab:params}), the Coulomb interaction will be dominated by the logarithmic part. This is also the case for a suspended MoS$_2$ monolayer ($d=6.14$~\AA, $\varepsilon=18.8$) \cite{Cheiwchanchamnangij:PRB2012}, with $r_0\approx 58$~\AA. We assume this configuration, $\varepsilon_1=\varepsilon_2=1$, throughout the rest of this work. From the variational approach in Ref. \cite{Prada:PRB15} the expressions for the exciton radii in $x$ and $y$ directions read $a_x=\sqrt{a_0r_0/[\mu_x^{-1}+(\lambda\mu_y)^{-1}]}$ and $a_y=\lambda a_x$,
where $a_0$ is the Bohr radius, and $\lambda=\left(\mu_x/\mu_y\right)^{1/3}$ measures the mass anisotropy. The binding energy $E_b$ in the same approximation reads
$E_b=-\frac{e^2}{r_0}\left\{\frac{3}{2}+\ln\left[(a_x+a_y)/8r_0\right]\right\}$.

\begin{table}[t]
\begin{tabular}{c| c c c c c c c c c c}
\hline
\hline
& $E_\mathrm{gap}$ & $E_b$ & $a_x$ & $a_y$ & $M_x$ & $M_y$ & $|v^{cv}|^2$ & $|\phi(0)|$ & $\tau_0$ & $\tau_{300 K}$\\
\hline
MoS$_2$  & 1.82 & 0.59 & 7.73 & 7.73 & 1.08 & 1.08 & 36.4 & 0.10 & 0.05 & 322\\
BP$_{1}$ & 1.84 & 0.59 & 10.8 & 5.10 & 0.36 & 5.02 & 53.4 & 0.11 & 0.03 & 249\\
BP$_{3}$ & 0.87 & 0.54 & 11.9 & 5.72 & 0.30 & 2.76 & 27.9 & 0.10 & 0.02 & \
1510\\
\hline
\hline
\end{tabular}
\caption{Exciton parameters for unstrained MoS$_2$ monolayers and BP mono- and trilayers, measured in electronvolts, Angstroms, electron masses, and picoseconds. $|v^{cv}|^2$ stands for $
\sum_i |v^{cv}_i|^2$. See footnote \cite{Note1} regarding the value of $E_\mathrm{gap}$ and $E_\mathrm{ex}$ for MoS$_2$.}
\label{tab:params}
\end{table}

Due to the presence of the electromagnetic environment, an exciton is merely a quasibound state of finite lifetime. Its main decay channel is through the emission of a photon of energy equal to that of the exciton $E(\vec Q)$ and of wavevector $\vec k=(Q_x,Q_y,k_z)$ for some out of plane $k_z$. Since the photon energy is $E=\hbar c |\vec k|$, this constraint can only be satisfied if $\hbar c|\vec Q|\lesssim E_\mathrm{ex}$, i.e. for small momentum excitons within the narrow light cone depicted in Fig. \ref{fig:bands}f. These `bright' excitons decay with a finite rate $\Gamma_{\vec Q}$ (see Appendix \ref{ap:decay} for a derivation and general  expressions). Around $\vec Q=0$, the decay rate reads
\begin{equation}\label{gammaQ}
\Gamma_{\vec Q\approx 0}=\frac{1}{\tau_0}=\frac{2\pi}{\hbar}\frac{\alpha}{E_\mathrm{ex}}|\phi(0)|^2\sum_{i={x,y}}|v^{cv}_i|^2,
\end{equation}
while $\Gamma_{\vec Q}=0$  for `dark' excitons outside the cone, within this particular decay channel. Here $\alpha\approx 1/137$ is the fine structure constant, $\phi(0)=\sqrt{2/(\pi a_x a_y)}$ is the exciton wavefunction \cite{Prada:PRB15} at $r=0$, and $v^{cv}_i$ are the valence-conduction dipole matrix elements $v^{cv}_{x,y}=\langle \psi_c(0)|\partial_{k_{x,y}} H(\vec k)|\psi_v(0)\rangle$, where $H(\vec k)$ denotes the tight-binding Bloch Hamiltonian and $\psi_{c,v}(0)$ are its single-particle eigenstates at either side of the (direct) gap. 

Typical intrinsic lifetimes $\tau_0$ around $\vec Q=0$ are very short, at around $\tau_0=30$ fs for monolayer BP and $\tau_0=100$ fs for monolayer MoS$_2$. It is known from experiments \cite{Korn:APL11} that the exciton lifetime dramatically increases with temperature, likely due to fast phonon-scattering of excitons into non-decaying `dark' states, such as those depicted in Fig. \ref{fig:bands}f. A simple argument based on instantaneous thermalization has been proposed \cite{Palummo:NL15} that, generalized to anisotropic exciton masses, yields the following lifetime for temperatures higher than $\sim 0.1$ K, 
\begin{equation}\label{Palummo}
\tau=\tau_0\frac{3}{2}k_B T \frac{\sqrt{M_xM_y}c^2}{E_\mathrm{ex}^2}. 
\end{equation}
This simple estimate predicts greatly enhanced $\tau\approx 249$ ps and $\tau\approx 525$ ps room-temperature lifetimes for BP and MoS$_2$ excitons, respectively, both within order-of-magnitude range of experimental results in pristine samples \cite{Korn:APL11,Amani:S15,Yang:LSA15,Surrente:PRB16}. (It should be noted that currently available experimental results for time-resolved exciton decay remain notoriously sample dependent, probably due to the effect of sample preparation, disorder, environmental screening, and the intrinsic complexity of out-of-equilibrium exciton dynamics.) Table \ref{tab:params} summarises  the above exciton properties for unstrained BP and MoS$_2$ monolayers. 

A generic strain field $\bm{\epsilon}(x,y)=\epsilon_{ij}(x,y)$ ($i,j=x,y,z$) can be efficiently incorporated into the hopping amplitudes of our tight-binding model. We denote by $t^{0}_{\alpha,\alpha'}$ the $\bm{\epsilon}=0$ hopping amplitudes between any two Wannier orbitals $\alpha,\alpha'$ sitting at positions $\vec{r}^0_\alpha$, $\vec{r}^0_{\alpha'}$, and connected by vector $\vec{r}^0_{\alpha\alpha'}=\vec{r}^0_{\alpha'}-\vec{r}^0_\alpha$. Under finite strain $\bm{\epsilon}$, hoppings are modified as $t_{\alpha\alpha'}=t^{0}_{\alpha\alpha'}\exp\left[-\beta_{\alpha\alpha'}(|\vec r_{\alpha\alpha'}|/|\vec r^0_{\alpha\alpha'}|-1)\right]$, where semi-phenomenological parameters $\beta_{\alpha\alpha'}=-d \ln t_{\alpha\alpha'}(r)/d \ln (r)|_{r=|\vec{r}^0_{\alpha\alpha'}|}$ are the dimensionless local electron-phonon couplings \cite{Suzuura:PRB02}, and $\vec r_{\alpha\alpha'}=\vec r^0_{\alpha\alpha'}+\bm{\epsilon}\cdot\vec r^0_{\alpha\alpha'}$ are the inter-site vectors modified by the strain tensor $\bm{\epsilon}$ at the bond. For simplicity we assume that $\beta_{\alpha\alpha'}$ depend solely on the $L^2$ angular momentum of the $\alpha$ and $\alpha'$ orbitals, not on their $L_z$ projections. Thus, BP has a single parameter which we take as $\beta_{pp}\approx 4.5$, while MoS$_2$ has three, $\beta_{pp}=3,\beta_{pd}=4,\beta_{dd}=5$ (the latter are consistent with the Wills-Harrison rule \cite{Harrison:99}). Due to a lack of accurate estimates of the above parameters in the literature, these values have once more been chosen here on the basis of ab-initio calculations, specifically by matching the direct-to-indirect gap transitions under strain in monolayers (at -4\% and 6.7\% uniaxial in BP \cite{Peng:PRB14}, and at 2-3\% biaxial in MoS$_2$ \cite{Feng:NP12,Wang:ADP14}). \edit{Appendix \ref{ap:benchmark} shows a comparison between the above theory and state-of-the-art ab-initio calculations for the exciton binding energy, both as a function of biaxial strain and number of layers.}

\begin{figure}[t]
\includegraphics[width=7.3cm]{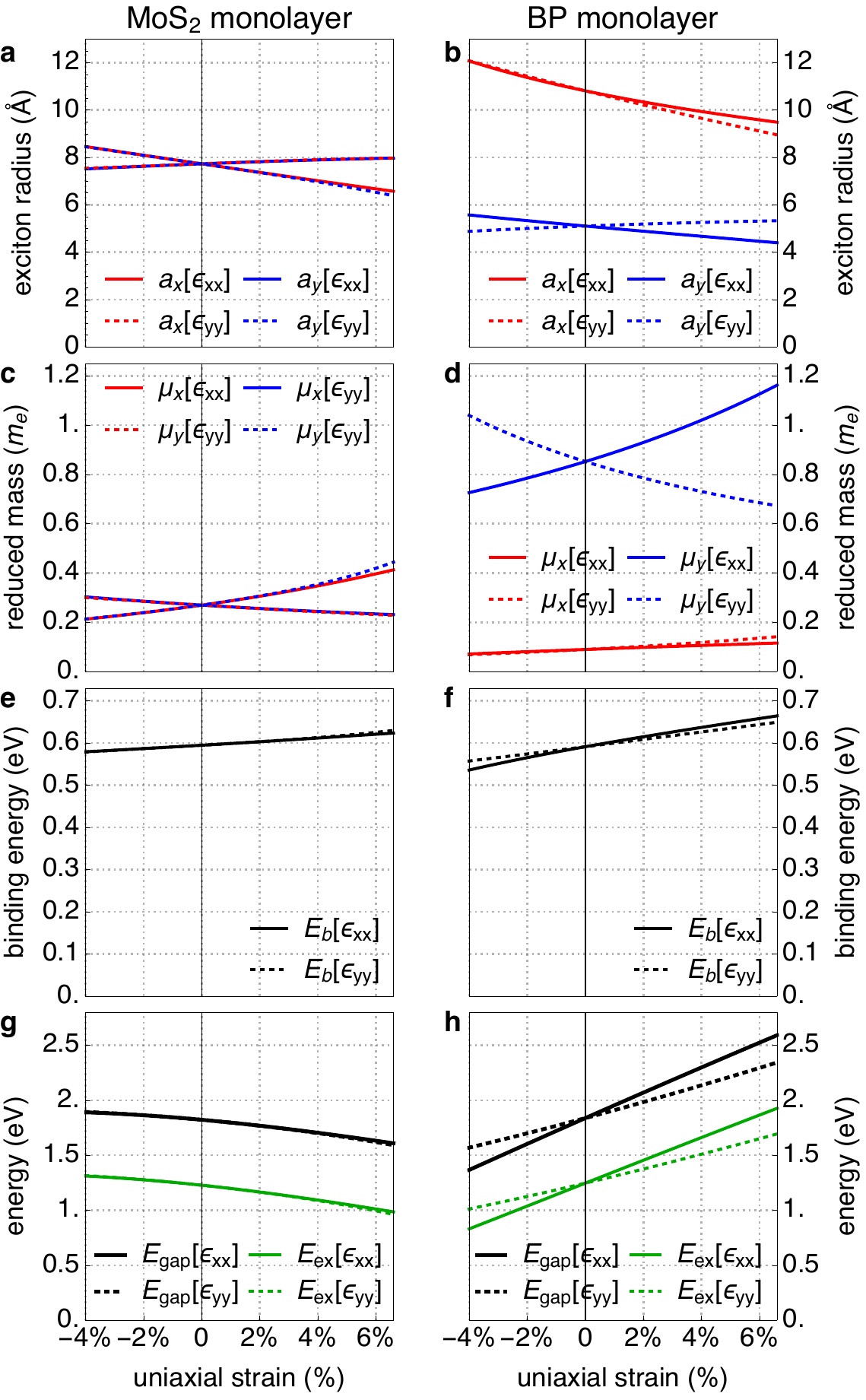}
\caption{Color online: Exciton properties as a function of uniaxial strain in the $x$ (armchair) and $y$ (zigzag) directions for MoS$_2$ (left column) and BP (right column) monolayers. Both materials remain in a direct gap regime within the chosen range of strains. (a,b) Exciton radii $a_{x,y}$. (c,d) Reduced exciton masses $\mu_{x,y}$. (e,f) Binding energy $E_b$. (g,h) Band gap $E_\mathrm{gap}$ and exciton energy $E_\mathrm{ex}$.}
\label{fig:exciton}
\end{figure}

\begin{figure}[t]
\includegraphics[width=7.3cm]{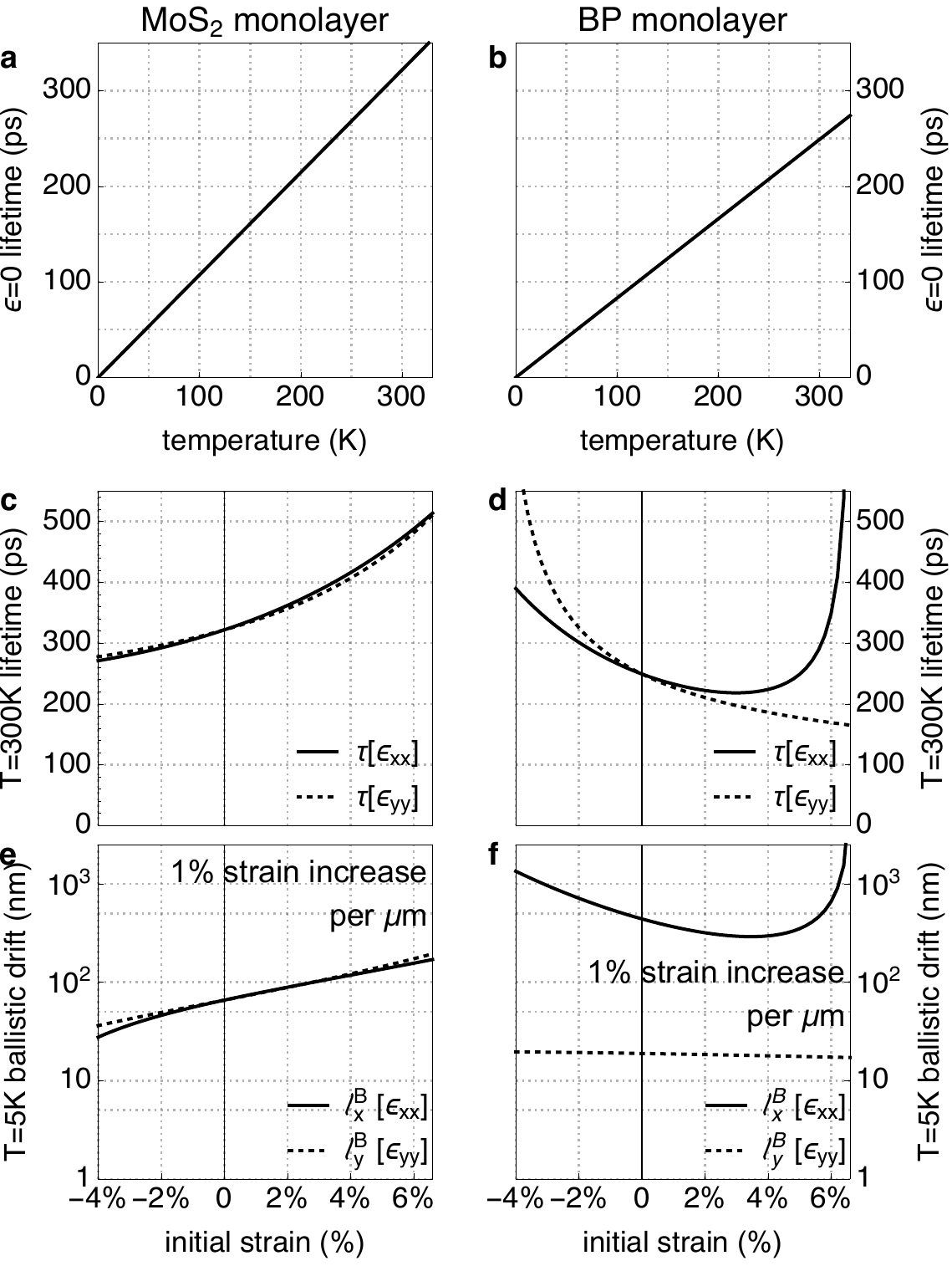}
\caption{(a,b) Temperature dependence of exciton lifetime $\tau$ without strain, and (c,d) lifetime versus uniaxial strain at room temperature, obtained from Eqs. (\ref{gammaQ},\,\ref{Palummo}). 
(e,f) Ballistic exciton funnel distances $\ell^B_{x,y}$ before recombination at $T=5$~K, as a function of initial strain $\epsilon_{ii}^0$, when subjected to a 1\% uniaxial strain increase per micrometer travelled. This measures the efficiency of the funnel effect along each uniaxial strain direction.}
\label{fig:dynamics}
\end{figure}

\section{Results and Discusion} 
We now characterize how the properties of excitons change under uniform uniaxial strains along armchair ($\epsilon_{xx}$) and zigzag ($\epsilon_{yy}$) directions on BP and MoS$_2$ monolayers. The corresponding strain tensors are diagonal, $\bm{\epsilon}_\mathrm{AC}=\epsilon_{xx}\mathrm{diag}(1, -\nu^\mathrm{AC}_y, -\nu^\mathrm{AC}_z)$ and $\bm{\epsilon}_\mathrm{ZZ}=\epsilon_{yy}\mathrm{diag}(-\nu^\mathrm{ZZ}_x, 1, -\nu^\mathrm{ZZ}_z)$. Poisson ratios $\nu$ depend on the effective elastic coefficients of the two materials, and are estimated to be $(\nu^\mathrm{AC}_y, \nu^\mathrm{AC}_z,\nu^\mathrm{ZZ}_x,\nu^\mathrm{ZZ}_z)\approx(0.7,-0.03,0.2,0.2)$ for BP monolayers \cite{Appalakondaiah:PRB12,Wei:APL14,Jiang:NC14,Peng:PRB14,Elahi:PRB15}, and $(\nu^\mathrm{AC}_y, \nu^\mathrm{AC}_z,\nu^\mathrm{ZZ}_x,\nu^\mathrm{ZZ}_z)\approx(0.25,0.0,0.25,0.0)$ for MoS$_2$ monolayers \cite{Yue:PLA12}. Although some uncertainty exists in these Poisson ratios, we observe that their precise values have little effect on the exciton properties under strain. 

The strain-dependence of exciton radii $a_{x,y}$, reduced masses $\mu_{x,y}$, binding energy $E_b$, band gap $E_\mathrm{gap}$ and exciton energy $E_\mathrm{ex}$  are presented in Fig. \ref{fig:exciton}, both for MoS$_2$ (left column) and BP monolayers (right column). In the former, although the gap remains direct, it is shifted slightly away from the K point as a result of the strain.  The most notable difference between the two materials is the strong anisotropy, apparent in the exciton shape and masses (panels a-d) and the opposite trend of the band gap with uniaxial strain: decreasing for MoS$_2$ (panel g) and increasing for monolayer BP (panel h, also true in multilayers). Due to the almost strain-independent binding energy $E_b$ in both cases (panels e,f), the exciton energy $E_\mathrm{ex}$, in green, also behaves this way under increasing uniaxial strain. In the case of biaxial strain the effect is even more pronounced (\edit{see Appendix \ref{ap:benchmark}}). Thus, an exciton generated on a sample with a finite strain gradient will be accelerated towards regions with higher tensile strain in monolayer MoS$_2$ (funnel effect), or away from said regions in few-layer BP (inverse funnel effect), as depicted in Figs. \ref{fig:funnel}.

\edit{The unusual sign of gap modulation with strain in BP ($\partial E_\mathrm{gap}/\partial \epsilon_{ii}>0$) as compared to transition metal dichalcogenides in general ($\partial E_\mathrm{gap}/\partial \epsilon_{ii}<0$), see Fig. \ref{fig:exciton}h, has  been demonstrated in optical absorption experiments \cite{Quereda:NL16} and ab-initio calculations \cite{Cakr:PRB14,Peng:PRB14}. It is ultimately a consequence of the puckered crystal structure of BP. The gap in this material, $E_\mathrm{gap}\approx 2t_2^{\parallel}+4t_1^{\parallel}>0$, is controlled directly by the partial cancellation between out-of-plane $t_2^{\parallel}$ and in-plane $t_1^{\parallel}$ hoppings,  which have opposite sign ($t_1^{\parallel}<0$ and $t_2^{\parallel}>0$, as defined in Fig. \ref{fig:parameters}). Due to the lattice puckering, tensile strains in the plane suppress $t_1^{\parallel}$, but increase $t_2^{\parallel}$ due to the positive out-of-plane Poisson ratio, leading to a rapid gap increase. }

The exciton lifetime $\tau$ for the two monolayers is shown in Figs. \ref{fig:dynamics}(a-d) versus temperature and strain. Its strain dependence is visibly stronger in BP than in MoS$_2$, even diverging at the strain-induced direct-to-indirect transitions ($\epsilon_{xx}=6.7\%$ and $\epsilon_{yy}=-4\%$), at which the BP monolayer valence band mass vanishes. Consider next the maximum distance an exciton may be funnelled across before it decays. 
We assume that the exciton does not dissociate under the acceleration (type-I funnel \cite{Feng:NP12}), which is the relevant regime for realistic strains in both these systems given their large binding energies. Take a perfectly ballistic sample with a linear spatial dependence of $E_\mathrm{ex}(\vec r)=\vec F\cdot \vec{r}$ produced by a strain gradient, $F_i=\partial_{\epsilon_{jk}}E_\mathrm{ex}(\bm\epsilon)\partial_{r_i}\epsilon_{jk}$. A semiclassical exciton subjected to the force $\vec F$ travels a distance $\ell^B_i=\frac{1}{2}\tau^2M^{-1}_iF_i$ before it decays \footnote{For simplicity, we assume the exciton masses $M_{x,y}$ to be constants, fixed to their initial values. A more accurate solution of the exciton motion using a position-dependent masses shows this is generally a rather accurate approximation. We also implicitly assume the strain gradient to be effectively adiabatic on the scale of the exciton radius.}. If the sample is disordered or temperature is high, the Drude scattering or phase coherence time $\tau_D$ due to defects or phonons may become shorter than the exciton's lifetime $\tau$. Its propagation then becomes diffusive before decaying, and the travelled distance is reduced to $\ell^D_i \approx \tau_D \tau M^{-1}_iF_i$ \cite{Bagaev:JL69,Shimizu:JOL06,Feng:NP12}. 
Figs. \ref{fig:dynamics}(e,f) show the ballistic funnel distances $\ell^B_i$ at $T=5K$ ($\tau\sim 4$ ps) traveled by an exciton generated at initial point $\vec r=(x_0,y_0)$ under a linear uniaxial strain profile $\epsilon_{xx}=\epsilon^0_{xx}+g (x-x_0)$ or $\epsilon_{yy}=\epsilon^0_{yy}+g (y-y_0)$. We consider a small strain gradient $g=$  1\% per $\mu$m, and plot $\ell^B_i$ as a function of initial strain $\epsilon^0_{ii}$. In a MoS$_2$ monolayer $\ell^B$ is isotropic and of the order of $\sim 70$ nm at zero initial strain, always towards increasingly strained regions. In BP monolayer, the ballistic (inverse) funnel distance is instead highly anisotropic, reaching $\sim 440$ nm along armchair and $\sim 20$ nm along zigzag directions, always away from strained regions. 
As a result, exciton flow becomes focused along the armchair direction in sufficiently ballistic samples, a phenomenon that may once more be exploited in various optoelectronic applications, as it will boost the device performance for specific orientations of electrodes or terminals, such as in the solar cell of Fig. \ref{fig:funnel}b. 

\edit{A key aspect for photocurrent generation \cite{Buscema:CSR15} in a funnel solar cell is the efficiency of exciton dissociation at the harvesting regions. This will critically depend on the contact properties, in particular the band alignment between BP and the semiconducting electrodes and the quality of the contact. The electrode materials and  configuration should be chosen so as to form a p-n junction at the contact that may  tear the exciton apart, converting its energy to electrical power with optimal efficiency. A number of recent works have been devoted to the properties of contacts to BP \cite{Pan:COM16}, with a focus on photovoltaic efficiency \cite{Dai:JPCL14,Deng:AN14,Buscema:NC14,Ganesan:APL16}. It has been predicted in particular that MoS${}_2$ \cite{Dai:JPCL14,Deng:AN14} or compressed BP itself \cite{Ganesan:APL16} could be ideal electrode materials for BP-based solar cells.}
 
The remarkable performance of the inverse funnel effect in BP monolayers is largely due to the small exciton mass $M_x$ along the armchair direction, see Table \ref{tab:params}, but also to the strong sensitivity of $E_\mathrm{gap}$ and $E_\mathrm{ex}$ with strain, Fig. \ref{fig:exciton}h. \edit{The strain modulation of the binding energy $E_b$ gives a relatively minor correction, so that more complex bound states that are formed at high excitation regimes, such as biexcitons \cite{Chaves:PRB16}, are expected to respond to strain gradients in a similar way as excitons, albeit possibly with reduced lifetimes at high densities \cite{Amani:S15}}. The efficient modulation of optoelectronic properties with strain, a hallmark property of this material, was recently showcased by optical absorption experiments in elastically rippled few-layer BP \cite{Quereda:NL16}. Increasing the number of BP layers, moreover, the inverse funnel performance is expected to improve even further.  As the gap is reduced, $E_\mathrm{ex}$ shifts down to energies with a far smaller photon density (the photon density of states is $\rho(E)=8\pi E^2/(hc)^3$), and the range of bright excitons shrinks. This produces a sharp increase of exciton lifetimes, see Table \ref{tab:params}. Moreover, while the strain sensitivity of the exciton energy $\partial_\epsilon E_\mathrm{ex}$ remains mostly unchanged, the averaged exciton mass decreases by up to $\sim40\%$, which conspires to increase the funnel distance even further \edit{as the number of layers increases (more details on multilayer funnelling  can be found in Appendix \ref{ap:multilayers}).} As an example, a ballistic three-layer BP sample is expected to reach values of $\ell^B_x$ in the tens of micrometers at $T=5$~K.   A real BP trilayer would obviously be in the diffusive funnel regime in this case, and additional decay channels may also have to be considered \cite{Thilagam:15}, but even with a $\tau_D \sim 1$ps, one would expect an $\ell^D_x$ of several micrometers. This renders few-layer BP a far more promising platform for exciton funnelling than MoS$_2$. 

To conclude, we have characterised the properties of Wannier excitons in few-layer BP and MoS$_2$ monolayers under strain. We have shown that the former presents strongly anisotropic exciton properties and a high sensitivity to strain. As a result we have demonstrated that few-layer BP should exhibit a remarkably strong anisotropic inverse funnel effect, which could be exploited for a number of optoelectronic technologies, such as high efficiency funnel solar cells.
While we have focused on the case of BP, our proposal of inverse exciton funnelling could potentially be realised in other, structurally similar compounds, such as group-IV monochalcogenides (e.g. GeSe) \cite{Gomes:PRB15,Hu:APL15}.
Some of these new materials (e.g. GeS or SnS), have multiple valleys and indirect gaps without strain, however. If they prove to be as strain-tuneable as BP, this feature could perhaps be turned into an advantage for exciton control. Assuming the indirect gap can be made direct under strain, one can envision strained, optically-active regions funnelling excitons towards dark, unstrained regions, which would result in enhanced lifetimes of accumulated excitons. Further opportunities to exploit the remarkable interplay between strain and exciton dynamics are also expected in twisted multilayers \cite{Wu:NL14}, and heterostructures combining several of these materials. Moir\'e patterns due to a lattice mismatch between layers are expected to give rise to gap modulations and spontaneous strain superlattices, phenomena already familiar from twisted graphene bilayers and graphene/boron nitride heterostructures \cite{Alden:PNAS13,Yankowitz:NP12,Woods:NP14,San-Jose:PRB14,Yankowitz:16}. Moir\'e patterns and strain superlattices could thus open the door to two-dimensional crystalline materials with built-in, spontaneous funnelling, without the need of externally induced strains.

\begin{acknowledgements}

We are grateful to A. Castellanos-G\'omez for illuminating discussions. We acknowledge financial support from MINECO (Spain) through the Ram\'on y Cajal program RYC-2013-14645 and RYC-2011-09345, and grant Nos. FIS2011-23713, FIS2013-47328-C2-1-P, FIS2014-58445-JIN, FIS2014-57432, and The “Mar\'ia de Maeztu” Programme for Units of Excellence in R$\&$D (MDM-2014-0377). Also from the Comunidad Aut\'onoma de Madrid (CAM) MAD2D-CM Program (S2013/MIT-3007), the European Commission under the Graphene Flagship, contract CNECTICT-604391, and the European Research Council, through grant No. 290846.

\end{acknowledgements}

\appendix

\section{Tight-Binding Models of Black Phosphorus and MoS$_2$}
\label{ap:TB}
\begin{figure}[t]
\includegraphics[width=\columnwidth]{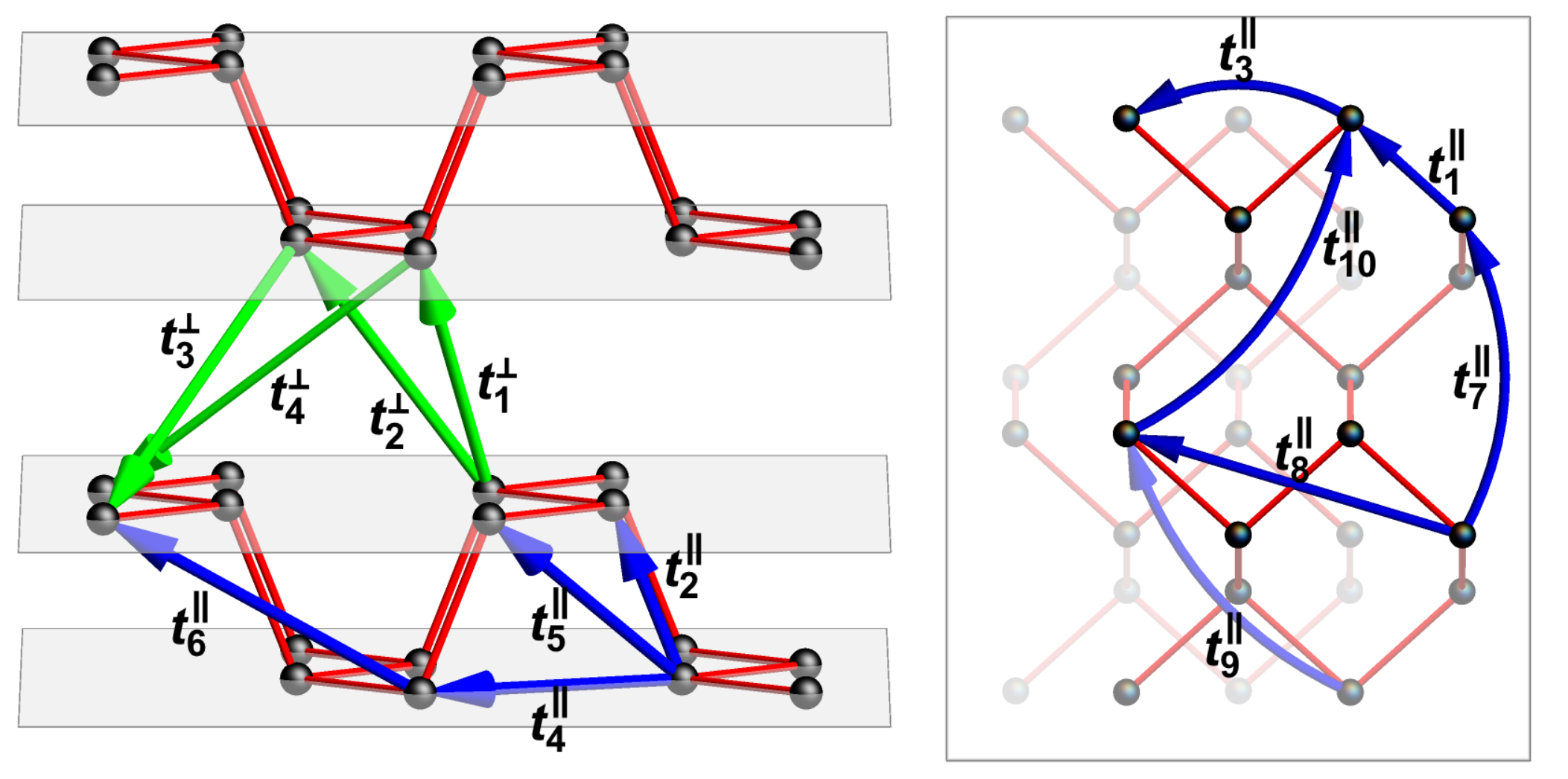}
\caption{Hopping amplitudes between $p_z$ orbitals in a BP multilayer. See Table \ref{tab:hoppingsBP} for their values.}
\label{fig:parameters}
\end{figure}

The computation of the exciton properties in the main text rely on the ability to obtain the gap, effective masses and dipole matrix elements $v^{cv}_{x,y}$ of the different materials under study, with and without strain. To this end, we employ a tight-binding model fitted to ab-initio calculations, and incorporate strains as discussed in the main text. For BP we used the model by Rudenko \emph{et al.} of Ref. \cite{Rudenko:PRB15}, which considers 14 hopping parameters between the $p_z$ orbital at each phosphorus atom, see Fig. \ref{fig:parameters}. Rudenko \emph{et al.} fit the hopping parameters to GW-LDA calculation without strain, with values reproduced in Table \ref{tab:hoppingsBP}.
The model includes hoppings between atoms with relative distance up to 5.49 \AA, and requires no crystal fields. It has been shown to accurately describe the bandstructure of BP from the monolayer to the bulk \cite{Rudenko:PRB15, Quereda:NL16}.

\begin{table}[t]
\begin{tabular}{|l|l|l|}
\hline
\hline
$t^{\parallel}_1 = -1.486$ eV & $t^{\parallel}_6 = 0.186$ eV  & $t^{\perp}_1 = 0.524$ eV\\
$t^{\parallel}_2 = 3.729$ eV  & $t^{\parallel}_7 = -0.063$ eV & $t^{\perp}_2 = 0.180$ eV\\
$t^{\parallel}_3 = -0.252$ eV & $t^{\parallel}_8 = 0.101$ eV  & $t^{\perp}_3 = -0.123$ eV\\
$t^{\parallel}_4 = -0.071$ eV & $t^{\parallel}_9 = -0.042$ eV & $t^{\perp}_4 = -0.168$ eV\\
$t^{\parallel}_5 = -0.019$ eV & $t^{\parallel}_{10} = 0.073$ eV &  \\
\hline
\hline
\end{tabular}
\caption{Values of the BP hopping parameters up to a distance of 5.49 \AA., obtained in Ref. \cite{Rudenko:PRB15} from a fit to GW-LDA results.}
\label{tab:hoppingsBP}
\end{table}

For MoS$_2$ monolayers we have used the model in Ref. \cite{Cappelluti:PRB13}, which includes all $p$ orbitals in sulfur atoms, and all d orbitals in molybdenum atoms. We have obtained the tight-binding parameters and crystal fields for this model by fitting to LDA bandstructure results (see footnote \cite{Note1}). The resulting values are shown in Table \ref{tab:hoppingsMoS2}. 

\begin{table}[t]
\begin{tabular}{l c r}
\hline
\hline
Crystal Fields & $\Delta_0$ & -1.094 eV\\
  & $\Delta_1$ & -0.05 eV\\
  & $\Delta_2$ & -1.512 eV\\
  & $\Delta_p$ & -3.560 eV\\
  & $\Delta_z$ & -6.886 eV\\
  &&\\
Intralayer Mo-S & $V_{pd\sigma}$ & 3.689 eV\\
  & $V_{pd\pi}$ & -1.241 eV\\
  &&\\
Intralayer Mo-Mo & $V_{dd\sigma}$ & -0.895 eV\\
  & $V_{dd\pi}$ & 0.252 eV\\
  & $V_{dd\delta}$ & 0.228 eV\\
  && \\
Intralayer S-S & $V_{pp\sigma}$ & 1.225 eV\\
  & $V_{pp\pi}$ & -0.467 eV\\
\hline
\hline
\end{tabular}
\caption{Tight-binding parameters employed in the MoS$_2$ model.}
\label{tab:hoppingsMoS2}
\end{table}

\section{Radiative Lifetime of excitons in Black Phosphorus}
\label{ap:decay}
Radiative recombination occurs when an exciton in its ground state $|\Psi_\mathrm{ex}(\vec Q)\rangle$, of energy $E_{ex}(\vec Q)$,  recombines with the consequent emission of a single photon in the state $|\gamma_{\vec k,\nu}\rangle = a^{\dagger}_\nu(\vec k)|0_\mathrm{em}\rangle$ with energy $\hbar\omega_{\vec k}=\hbar c |\vec k|$. The operator $a^{\dagger}_\nu(\vec k)$ creates a photon with momentum $\vec k$ polarized along the vector $\vec{e}_\nu$. In this section we present a detailed derivation of general expressions for the decay rate of an exciton within this single-photon channel using a generic two-dimensional tight-binding description for the system, and the general description of excitons of Ref. \cite{Prada:PRB15}.

\subsection{Model}

Assume a 2D system with a generic Bloch Hamiltonian $H(\vec k)$ obtained e.g. from a tight-binding model. For simplicity, we further assume the system has a direct gap at the $\Gamma$ point (the final expressions will still be valid for expansions around a different point, as long as we measure momenta from that point), so that at small $|\vec k|$ we may expand
\[
H(\vec k)= H_0+k_i H^i_1+ \frac{1}{2} k_i k_j H_2^{ij}+\mathcal{O}(k^3),
\]
with $H_1^i=\partial_{k_i}H$ and $H_2^{ij}=\partial_{k_i}\partial_{k_j}H$, evaluated at $\vec k=0$. In the presence of an electromagnetic environment, the minimal coupling enters as $H(\vec k -\frac{e}{\hbar}\vec A)$, where $\vec A$ is the electromagnetic field. The expanded Hamiltonian becomes, to first order in $\vec A$
\[
H(\vec k-\frac{e}{\hbar}\vec A)= H(\vec k) +W_\mathrm{em}  +\mathcal{O}(A^2),
\]
with the electromagnetic vertex defined as
\[
W_\mathrm{em} = - \frac{e}{\hbar}A_i(\vec r) V^i
\]
and
\begin{equation}\label{V}
V^i(\vec k)=H_1^i + k^j H_2^{ij}.
\end{equation}

The second-quantization version of $H(\vec k)$ is obtained as usual,
\[
H(\vec k) = \sum_{\vec k}^\mathrm{2D} c^\dagger_s(\vec k) \left[H(\vec k)\right]_{ss'}c_{s'}(\vec k),
\]
where $s, s'$ are indices in a basis in the unit cell, and are implicitly summed over. 
The electromagnetic vertex is similarly expressed as
\[
W_\mathrm{em}=-\frac{e}{\hbar}\int d^2r A_i(\vec r) \psi_s(\vec r) \left[V^i\right]_{ss'} \psi_{s'}(\vec r),
\]
where the $k_j$ term in $V^i$ above gives rise to a $-i\partial_{x_j}\psi_{s'}(\vec r)$.

The Hamiltonian of the electromagnetic environment, derived from quantizing the electromagnetic action $\mathcal{S}=\frac{1}{4}\int d^4x F_{\mu\nu}F^{\mu \nu}$, can be written as
\[
H_\mathrm{em}=\sum_{\vec k,\nu=\pm1}^\mathrm{3D}\hbar \omega
_{\vec k}\left[a_\nu^{\dagger}(\vec k)a_\nu(\vec k)+\frac{1}{2}\right],
\]
where $\omega_{\vec k}=c|\vec k|$, and
\[
\vec A(\vec r)=\sum_{\vec k, \nu=\pm 1}^\mathrm{3D}\sqrt{\frac{\hbar}{2\omega_{\vec k} \Omega\epsilon_0}}\left[\vec e_\nu a_\nu(\vec k)e^{i\vec k\vec r}+\mathrm{h.c.}\right].
\]
In the above equations, $\Omega$ is the total volume of the system, and $\nu=\pm 1$ are the two possible polarizations of the photon field, so that $\vec k\cdot\vec e_\nu=0$. If $\vec k=k \hat z$, for example, then $\vec e_\pm=(\mp\hat x - i \hat y)/\sqrt{2}$ for a basis with circular polarization. This form of $\vec A$ corresponds to the Coulomb (or transverse) gauge $\vec\nabla\cdot \vec A=0$, for which $\vec A$ oscillates in the plane perpendicular to the propagation direction of the photon.

The complete vertex in second quantization then reads
\begin{eqnarray}\label{Wem}
W_\mathrm{em}&=&-\frac{e}{\hbar}\sum_{\vec k\nu}^\mathrm{3D}\sum_{\vec k'}^\mathrm{2D}\sqrt{\frac{\hbar}{2\omega_{\vec k}\Omega\epsilon_0}}e^i_\nu a_\nu(\vec k)\\
&&\times c_s^\dagger(\vec k'+\frac{\vec k_\parallel}{2})V^i_{ss'}(\vec k')c_{s'}(\vec k'-\frac{\vec k_\parallel}{2})+\mathrm{h.c.}\nonumber
\end{eqnarray}
The notation $\vec k_\parallel$ above stands for the photon wavevector within the sample plane. Note the implicit summation over $i$ (dot product of gauge field and fermionic current).

\subsection{Exciton ground state}

We consider now an exciton, i.e. an electron-hole pair bound by Coulomb interaction and with energy within the gap. As discussed in Ref. \cite{Prada:PRB15}, the wavefunction of the pair may be obtained from a Schr\"odinger equation in the electron-hole relative coordinate $\vec r$, which has a reduced mass tensor $(\mu_{ij})^{-1}=(m^{c}_{ij})^{-1}+(m^{v}_{ij})^{-1}$, in terms of the effective mass tensors of the electron and the hole at the conduction and valence bands. The solution for the wavefunction is $\phi(\vec r)$. An exciton with a total momentum $\vec Q$ has energy $E_\mathrm{ex}(\vec Q)=E_\mathrm{gap}(\vec Q)-E_b(\vec Q)$, see Fig. \ref{fig:bands}f in the main text, where $E_\mathrm{gap}(\vec Q)$ is the gap of $H(\vec Q)$ and $E_b(\vec Q)$ is the exciton binding energy. The particle-hole state is written as
\[
|\Psi_\mathrm{ex}(\vec Q)\rangle=\int d^2R\, d^2r\, \frac{e^{-i\vec Q\vec R}}{\sqrt{S}}\phi(\vec r)\psi^\dagger_c(\vec R-\vec r/2)\psi_v(\vec R+\vec r/2)|0\rangle,
\]
where $S$ is the surface of the system, $\vec R$ is the center-of-mass coordinate and $|0\rangle$ is the electronic system's ground state. We may Fourier transform the above using $c^\dagger(\vec k)=\frac{1}{\sqrt{S}}\int d^2r \, e^{i\vec k\vec r}\psi^\dagger(\vec r)$ and its converse $\psi^\dagger(\vec r)=\frac{1}{\sqrt{S}}\sum_{\vec k}^\mathrm{2D}e^{-i\vec k\vec r}c^\dagger(\vec k)$. This gives, for the exciton state at momentum $\vec Q$,
\[
|\Psi_\mathrm{ex}(\vec Q)\rangle=\frac{1}{\sqrt{S}}\sum_{\vec k}^\mathrm{2D}\int d^2r\, e^{i\vec k \vec r}\phi(\vec r) c^\dagger_c(\vec k+\frac{\vec Q}{2})c_v(\vec k-\frac{\vec Q}{2})|0\rangle.
\]

\subsection{Exciton decay}

We wish to find the relaxation rate of the $|\Psi_\mathrm{ex}(\vec Q)\rangle$ exciton due to its coupling $W_\mathrm{em}$ to the electromagnetic environment. According to the Fermi golden rule, this rate is
\[
\Gamma_{\vec Q}=\frac{2\pi}{\hbar}\sum_{\vec k \nu}^\mathrm{3D}\left|\langle 0; \gamma_{\vec k,\nu}|W_\mathrm{em}|\Psi_\mathrm{ex}(\vec Q);0_\mathrm{em}\rangle\right|^2 \delta(\hbar\omega_{\vec k}-E_\mathrm{ex}(\vec Q)).
\]
Note that $|\gamma_{\vec k,\nu}\rangle = a^{\dagger}_\nu(\vec k)|0_\mathrm{em}\rangle$ is a single photon state, and $|0_\mathrm{em}\rangle$ is the electromagnetic vacuum. 
We insert the form of $W_\mathrm{em}$, Eq. (\ref{Wem}), to get,
\begin{eqnarray}\label{Gamma1}
\Gamma_{\vec Q}&=&\frac{2\pi}{\hbar}\frac{e^2}{\hbar^2}\frac{\hbar}{2\Omega\epsilon_0}
\sum_{\vec k, \nu}^\mathrm{3D}\frac{\left|\langle 0_\mathrm{em}|a_\nu(\vec k)a^\dagger_\nu(\vec k)|0_\mathrm{em}\rangle\right|^2}{\omega_{\vec k}} \nonumber \\
&&\times\left|\langle 0|\vec e^{i*}_\nu\cdot \vec V|\Psi_\mathrm{ex}(\vec Q)\rangle\right|^2 \delta(\hbar\omega_{\vec k}-E_\mathrm{ex}(\vec Q))\nonumber\\
&=&\frac{2\pi}{\hbar}\frac{e^2}{\hbar^2}\frac{\hbar^2}{2\Omega E_\mathrm{ex}\epsilon_0}
\sum_{\vec k, \nu}^\mathrm{3D}\left| \langle 0|\vec e^{i*}_\nu\cdot \vec V_{\vec k_\parallel}^\dagger|\Psi_\mathrm{ex}(\vec Q)\rangle\right|^2\nonumber\\
&&\times\delta(\hbar\omega_{\vec k}-E_\mathrm{ex}(\vec Q)),
\end{eqnarray}
where we define $\vec V_{\vec k_\parallel}=(V^x_{\vec k_\parallel},V^y_{\vec k_\parallel},0)$ and
\begin{equation}\label{V2}
V^i_{\vec k_\parallel}=\sum_{\vec k'}^\mathrm{2D}c_s^\dagger(\vec k'+\frac{\vec k_\parallel}{2})\left(v_i^{ss'}+{k'}^j w_{ij}^{ss'}\right)c_{s'}(\vec k'-\frac{\vec k_\parallel}{2}).
\end{equation}
The matrix element in Eq. (\ref{Gamma1}) produces the Kronecker constraint $\delta_{\vec k_\parallel-\vec Q}\delta_{s'v}\delta_{sc}$, and finally gives
\begin{eqnarray*}
\langle 0|V^i_{\vec k_\parallel}|\Psi_\mathrm{ex}(\vec Q)\rangle &=&
\frac{\delta_{\vec k_\parallel-\vec Q}}{\sqrt{S}}\int d^2r\sum_{\vec k'} e^{i\vec k'\vec r}V^{i*}_{cv}(\vec k')\phi(r)
\nonumber\\
&=&\sqrt{S}\delta_{\vec k_\parallel-\vec Q}\left(v_{i}^{cv*}\phi(0)-iw^{cv*}_{ij}\partial_{x_j}\phi(0)\right) \nonumber\\
&\equiv&\sqrt{S}\delta_{\vec k_\parallel-\vec Q} \mathcal{F}^{i*},
\end{eqnarray*}
where the star denotes complex conjugation and the parenthesis in the second line is denoted by $\mathcal{F}^{i}$ for brevity. From Eq. \eqref{V}, the dipole matrix elements $v^{cv}_i$ and $w^{cv}_{ij}$ above are defined as first and second order derivatives of $H(\vec k)$  around the gap at $\vec k=0$, respectively,
\begin{eqnarray*}
v^{cv}_i&=&\langle \psi_c(0)|\partial_{k_i}H|\psi_v(0)\rangle,\\
w^{cv}_{i,j}&=&\langle \psi_c(0)|\partial_{k_i}\partial_{k_j}H|\psi_v(0)\rangle.
\end{eqnarray*}
We have also used $\frac{1}{S}\sum_{\vec k}^\mathrm{2D}e^{-i\vec k\vec r}=\delta(\vec r)$. We finally eliminate the sum over the photon's $\vec k_\parallel$  with the above $\delta_{\vec k_\parallel-\vec Q}$ and obtain
\begin{equation*}
\Gamma_{\vec Q}=\frac{2\pi}{\hbar}\frac{e^2}{\hbar^2}\frac{\hbar^2 S}{2\Omega E_\mathrm{ex}\epsilon_0}
\sum_{k_z\nu}|e^{i}_\nu\mathcal{F}^i|^2\delta(\hbar c\sqrt{Q^2+k_z^2}-E_\mathrm{ex}(\vec Q)).
\end{equation*}
We use the energy constraint to perform the sum $\sum_{k_z}=\frac{L_z}{2\pi}\int dk_z$ (where $L_z S=\Omega$), taking into account the appropriate Jacobian
\[
\delta(\hbar c\sqrt{Q^2+k_z^2}-E_\mathrm{ex}(\vec Q))=\frac{E_\mathrm{ex}(\vec Q)}{(\hbar c)^2k_x^{(0)}}\delta(k_z-k_z^{(0)}),
\]
with $k_z^{(0)}$ defined by $\hbar c\sqrt{Q^2+{k_z^{(0)}}^2}=E_\mathrm{ex}(\vec Q)$. 

Finally,
\begin{eqnarray*}
\Gamma_{\vec Q}&=&\frac{e^2}{\hbar}\frac{1}{2 \epsilon_0}\frac{1}{(\hbar c)^2k_z^{(0)}}
\sum_{i,\nu}|e^{i}_\nu \mathcal{F}^i|^2\\
&=&\frac{1}{\hbar}\frac{2\pi\alpha}{\sqrt{E_\mathrm{ex}(\vec Q)^2-(\hbar c Q)^2}}\\
&&\times\sum_{i,\nu}\left|e^{i}_\nu\left[v_{i}^{cv}\phi(0)+iw^{cv}_{ij}\partial_{x_j}\phi(0)\right]\right|^2.
\end{eqnarray*}

Recall that $\vec e_\nu$ (with its dual $\vec e^*_\nu$) form a two-dimensional and $\vec Q$-dependent orthonormal basis of the plane orthogonal to the photon's $\vec k$, namely $\vec k=(Q_x,Q_y,k_z^{(0)})$ in this case. Note also that $Q=|\vec Q|<E_\mathrm{ex}/\hbar c$ above, otherwise $\Gamma_{\vec Q}=0$. If $Q=0$ (exciton ground state), the photon has only $z$ momentum, and $\vec e_\nu$ can be chosen as $\hat x,\hat y$, i.e $e_\nu^i=\delta_{\nu,i}$. We then recover the expression given in the main text,
\begin{eqnarray}\label{Gamma}
\Gamma_{0}&=&\frac{1}{\hbar}\frac{2\pi\alpha}{E_\mathrm{ex}}\sum_{i}\left|v_{i}^{cv}\phi(0)+iw^{cv}_{ij}\partial_{x_j}\phi(0)\right|^2.
\end{eqnarray}

Note that the $w^{cv}\phi'(0)$ term can be neglected if $\phi(\vec r)$ is an even (differentiable) function of position. Likewise, the $v^{cv}\phi(0)$ term vanishes for an odd $\phi(\vec r)$, which is expected of an excited state for the exciton (not considered in this work).
Note also that the above derivation relies heavily on momentum conservation laws, which apply only if the sample size is much larger than the photon wavelength $\lambda=2\pi/k=hc/E_\mathrm{ex}$. In the opposite limit, the decay rate is expected to be linear in sample area, and the photon is emitted isotropically.

\section{Theory benchmarking}
\label{ap:benchmark}

\begin{figure}
\includegraphics[width=8 cm]{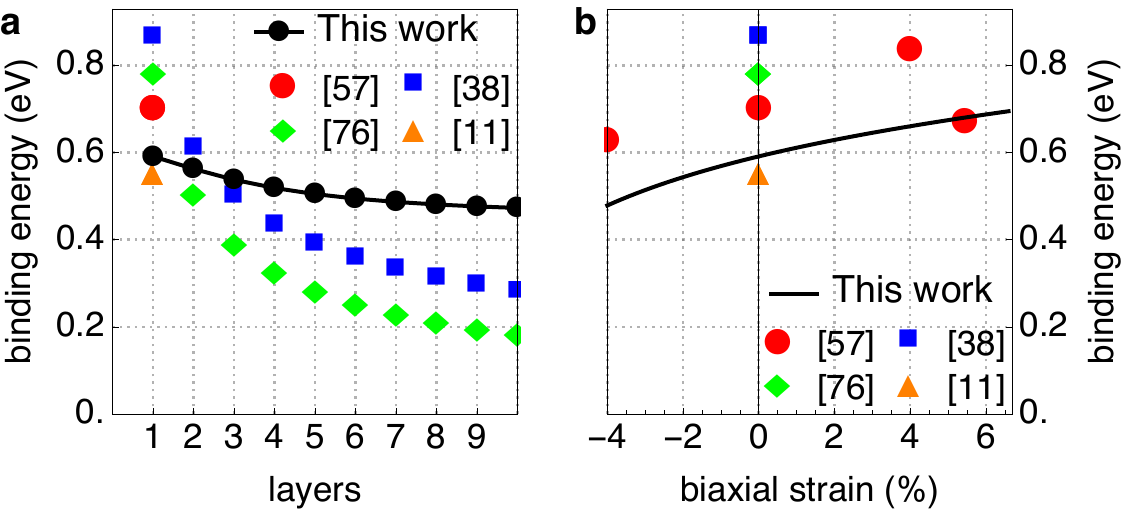}
\caption{\edit{(a) Exciton binding energy in unstrained BP as a function of number of layers. (b) Exciton binding energy in a BP monolayer as a function of biaxial strain. Coloured symbols correspond to the indicated references.}}
\label{fig:benchmark}
\end{figure}

\edit{In this section we assess the accuracy of our theory for the exciton binding energy $E_b$, by comparing to published ab-initio results based on the Bethe-Salpeter equation, see Refs. \cite{Cakr:PRB14,Tran:PRB14,Tran:2M15,Ganesan:APL16}. Figure \ref{fig:benchmark}a shows the comparison of $E_b$ in unstrained BP multilayers for increasing number of layers $n$. We find good agreement up to around $n=3$ layers, with our theory well within the dispersion of published predictions. Beyond $n=3$, the two-dimensional Keldysh potential employed in this work leads to an overestimation of the binding energy. The reason is that when the multilayer thickness exceeds the vertical exciton radius, a proper three-dimensional calculation of $E_b$ becomes necessary, which in turn leads to smaller binding energies, properly captured by the ab-initio results. This is a generic effect, whereby an increased dimensionality leads to reduced binding energies from confining potentials \cite{Zaslow:AJOP67}.}

\edit{Figure \ref{fig:benchmark}b shows a second comparison of $E_b$ in BP monolayers as a function of biaxial strain. We once more find reasonable agreement between our theory and ab-initio calculation. Note in particular that the trend of increasing $E_b$ with tensile strain is correctly captured. We thus conclude that our analytical description of exciton properties is quite accurate for BP multilayers up to $n=3$ within a wide range of realistic deformations.}

\section{Inverse funnel effect in BP multilayers}
\label{ap:multilayers}

\begin{figure}[t]
\includegraphics[width=8 cm]{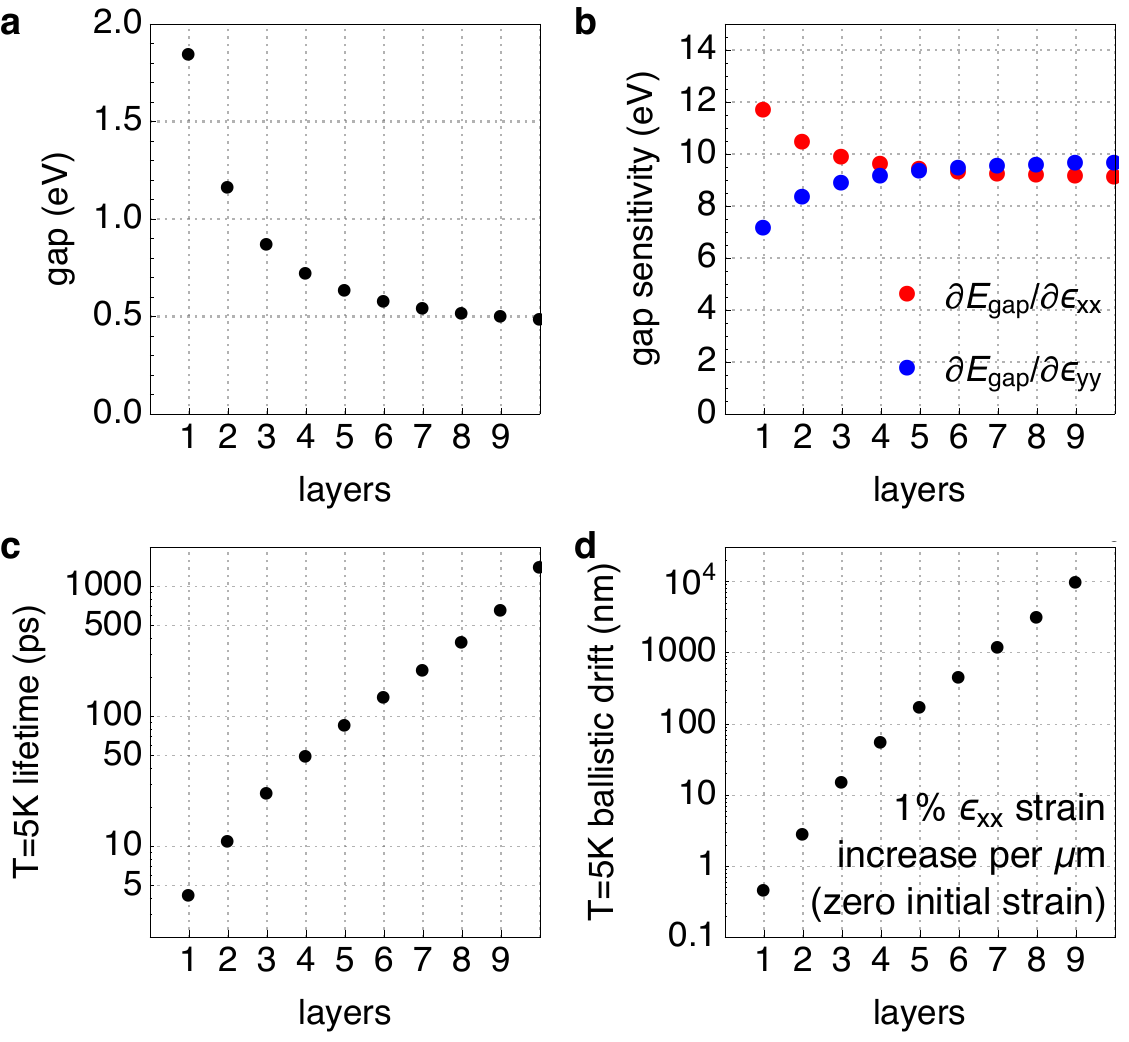}
\caption{\edit{(a) Direct bandgap, (b) bandgap change as a function of strain $\partial E_\mathrm{gap}/\partial\epsilon_{ii}$, (c) exciton lifetime at $T=5K$, and (d) ballistic drift length as a function of number of layers in a BP multilayer.}}
\label{fig:multilayer}
\end{figure}

\edit{The theory and simulations presented in the main text focus on the case of BP and MoS${}_2$ monolayers. While for MoS${}_2$, multilayers are less useful for optoelectronics, given their indirect gap, this limitation does not apply to BP multilayers, whose gap is direct irrespective of the number of layers $n$. Indeed, as anticipated in the main text, we expect the inverse funnel effect in multilayer BP to strongly outperform that of BP monolayers. The main reason is a strong enhancement of exciton lifetime as $n$ increases, which is a result of the decreasing bandgap (see Fig. \ref{fig:multilayer}a), and the correspondingly suppressed density of photon densities at smaller energies, $\rho(E)\sim E^2$. The sensitivity of the gap as a function of strain $\partial E_\mathrm{gap}/\partial\epsilon_{ii}$ is not dramatically affected by $n$, see Fig. \ref{fig:multilayer}b.
Our theory thus predicts a strong increase for the BP exciton lifetime as $n$ grows, see Fig. \ref{fig:multilayer}c. Similarly, the corresponding ballistic drift length, Fig. \ref{fig:multilayer}d, is enhanced into the tens of micrometers for ten layers.}

\edit{These multilayer predictions should be taken with caution, however. The theory, as presented, is strictly valid for excitons confined to two dimensions. When the multilayer thickness exceeds the vertical exciton diameter, the theory ceases to apply, strictly speaking. This happens already for a BP trilayer. The inclusion of the vertical dimension into the theory, however, is expected to only modify the results for the exciton binding energy, see Appendix \ref{ap:benchmark}, which is not essential to the funnel effect, as discussed in the main text. Indeed, the main driver for the efficient inverse funnel mechanism is the modulation of the single-particle bandgap with strain, Fig. \ref{fig:multilayer}b, which is correctly captured by our theory. Therefore, we expect the results in Fig. \ref{fig:multilayer}(c,d) to remain qualitatively correct in a more general three-dimensional theory.}

\edit{A second limitation, however, is not so straightforward to generalise. The only exciton decay channel considered here is radiative decay. Our theory does not include other non-radiative decay channels, such as Auger scattering, that while relatively unimportant for monolayers, should dominate exciton decay for thicker multilayers. Therefore, the results of Fig. \ref{fig:multilayer} should rather be interpreted as an upper bound, absent a more complete description of all exciton decay channels. }

\edit{One last consideration that should be kept in mind is the problem of inducing strain gradients in multilayers. As $n$ grows, the material becomes stiffer, which may challenge efforts to create a given strain gradient. Moreover, the appearance of interlayer shear becomes a possibility (not considered here) which would modulate the interlayer coupling spatially, adding considerable complexity to the problem. A complete study of the multilayer elasticity problem is far beyond the scope of the present paper, but should be carefully considered in multilayer experiments.}

\edit{Despite the above considerations, we anticipate that the optimal number of layers for the inverse funnel effect is probably greater than one, but not much greater than eight, for which the bandgap begins to saturate to its bulk value. The performance for the optimal BP multilayer is expected, in any case, to greatly exceed the already remarkable results predicted here for the BP monolayer.}

\bibliography{biblio}

\begin{thebibliography}{80}%
\makeatletter
\providecommand \@ifxundefined [1]{%
 \@ifx{#1\undefined}
}%
\providecommand \@ifnum [1]{%
 \ifnum #1\expandafter \@firstoftwo
 \else \expandafter \@secondoftwo
 \fi
}%
\providecommand \@ifx [1]{%
 \ifx #1\expandafter \@firstoftwo
 \else \expandafter \@secondoftwo
 \fi
}%
\providecommand \natexlab [1]{#1}%
\providecommand \enquote  [1]{``#1''}%
\providecommand \bibnamefont  [1]{#1}%
\providecommand \bibfnamefont [1]{#1}%
\providecommand \citenamefont [1]{#1}%
\providecommand \href@noop [0]{\@secondoftwo}%
\providecommand \href [0]{\begingroup \@sanitize@url \@href}%
\providecommand \@href[1]{\@@startlink{#1}\@@href}%
\providecommand \@@href[1]{\endgroup#1\@@endlink}%
\providecommand \@sanitize@url [0]{\catcode `\\12\catcode `\$12\catcode
  `\&12\catcode `\#12\catcode `\^12\catcode `\_12\catcode `\%12\relax}%
\providecommand \@@startlink[1]{}%
\providecommand \@@endlink[0]{}%
\providecommand \url  [0]{\begingroup\@sanitize@url \@url }%
\providecommand \@url [1]{\endgroup\@href {#1}{\urlprefix }}%
\providecommand \urlprefix  [0]{URL }%
\providecommand \Eprint [0]{\href }%
\providecommand \doibase [0]{http://dx.doi.org/}%
\providecommand \selectlanguage [0]{\@gobble}%
\providecommand \bibinfo  [0]{\@secondoftwo}%
\providecommand \bibfield  [0]{\@secondoftwo}%
\providecommand \translation [1]{[#1]}%
\providecommand \BibitemOpen [0]{}%
\providecommand \bibitemStop [0]{}%
\providecommand \bibitemNoStop [0]{.\EOS\space}%
\providecommand \EOS [0]{\spacefactor3000\relax}%
\providecommand \BibitemShut  [1]{\csname bibitem#1\endcsname}%
\let\auto@bib@innerbib\@empty
\bibitem [{\citenamefont {Feng}\ \emph {et~al.}(2012)\citenamefont {Feng},
  \citenamefont {Qian}, \citenamefont {Huang},\ and\ \citenamefont
  {Li}}]{Feng:NP12}%
  \BibitemOpen
  \bibfield  {author} {\bibinfo {author} {\bibfnamefont {Ji}~\bibnamefont
  {Feng}}, \bibinfo {author} {\bibfnamefont {Xiaofeng}\ \bibnamefont {Qian}},
  \bibinfo {author} {\bibfnamefont {Cheng-Wei}\ \bibnamefont {Huang}}, \ and\
  \bibinfo {author} {\bibfnamefont {Ju}~\bibnamefont {Li}},\ }\bibfield
  {title} {\enquote {\bibinfo {title} {Strain-engineered artificial atom as a
  broad-spectrum solar energy funnel},}\ }\href@noop {} {\bibfield  {journal}
  {\bibinfo  {journal} {Nat. Photon.}\ }\textbf {\bibinfo {volume} {6}},\
  \bibinfo {pages} {866--872} (\bibinfo {year} {2012})}\BibitemShut {NoStop}%
\bibitem [{\citenamefont {Li}\ \emph {et~al.}(2014{\natexlab{a}})\citenamefont
  {Li}, \citenamefont {Yu}, \citenamefont {Ye}, \citenamefont {Ge},
  \citenamefont {Ou}, \citenamefont {Wu}, \citenamefont {Feng}, \citenamefont
  {Chen},\ and\ \citenamefont {Zhang}}]{Li:NN2014}%
  \BibitemOpen
  \bibfield  {author} {\bibinfo {author} {\bibfnamefont {Likai}\ \bibnamefont
  {Li}}, \bibinfo {author} {\bibfnamefont {Yijun}\ \bibnamefont {Yu}}, \bibinfo
  {author} {\bibfnamefont {Guo~Jun}\ \bibnamefont {Ye}}, \bibinfo {author}
  {\bibfnamefont {Qingqin}\ \bibnamefont {Ge}}, \bibinfo {author}
  {\bibfnamefont {Xuedong}\ \bibnamefont {Ou}}, \bibinfo {author}
  {\bibfnamefont {Hua}\ \bibnamefont {Wu}}, \bibinfo {author} {\bibfnamefont
  {Donglai}\ \bibnamefont {Feng}}, \bibinfo {author} {\bibfnamefont {Xian~Hui}\
  \bibnamefont {Chen}}, \ and\ \bibinfo {author} {\bibfnamefont {Yuanbo}\
  \bibnamefont {Zhang}},\ }\bibfield  {title} {\enquote {\bibinfo {title}
  {Black phosphorus field-effect transistors},}\ }\href@noop {} {\bibfield
  {journal} {\bibinfo  {journal} {Nat. Nano.}\ }\textbf {\bibinfo {volume}
  {9}},\ \bibinfo {pages} {372--377} (\bibinfo {year}
  {2014}{\natexlab{a}})}\BibitemShut {NoStop}%
\bibitem [{\citenamefont {Koenig}\ \emph {et~al.}(2014)\citenamefont {Koenig},
  \citenamefont {Doganov}, \citenamefont {Schmidt}, \citenamefont {Neto},\ and\
  \citenamefont {Oezyilmaz}}]{Koenig:APL2014}%
  \BibitemOpen
  \bibfield  {author} {\bibinfo {author} {\bibfnamefont {Steven~P}\
  \bibnamefont {Koenig}}, \bibinfo {author} {\bibfnamefont {Rostislav~A}\
  \bibnamefont {Doganov}}, \bibinfo {author} {\bibfnamefont {Hennrik}\
  \bibnamefont {Schmidt}}, \bibinfo {author} {\bibfnamefont {AH~Castro}\
  \bibnamefont {Neto}}, \ and\ \bibinfo {author} {\bibfnamefont {Barbaros}\
  \bibnamefont {Oezyilmaz}},\ }\bibfield  {title} {\enquote {\bibinfo {title}
  {Electric field effect in ultrathin black phosphorus},}\ }\href@noop {}
  {\bibfield  {journal} {\bibinfo  {journal} {Appl. Phys. Lett.}\ }\textbf
  {\bibinfo {volume} {104}},\ \bibinfo {pages} {103106} (\bibinfo {year}
  {2014})}\BibitemShut {NoStop}%
\bibitem [{\citenamefont {Liu}\ \emph {et~al.}(2014)\citenamefont {Liu},
  \citenamefont {Neal}, \citenamefont {Zhu}, \citenamefont {Luo}, \citenamefont
  {Xu}, \citenamefont {Tom{\'a}nek},\ and\ \citenamefont {Ye}}]{Liu:ACS2014}%
  \BibitemOpen
  \bibfield  {author} {\bibinfo {author} {\bibfnamefont {Han}\ \bibnamefont
  {Liu}}, \bibinfo {author} {\bibfnamefont {Adam~T}\ \bibnamefont {Neal}},
  \bibinfo {author} {\bibfnamefont {Zhen}\ \bibnamefont {Zhu}}, \bibinfo
  {author} {\bibfnamefont {Zhe}\ \bibnamefont {Luo}}, \bibinfo {author}
  {\bibfnamefont {Xianfan}\ \bibnamefont {Xu}}, \bibinfo {author}
  {\bibfnamefont {David}\ \bibnamefont {Tom{\'a}nek}}, \ and\ \bibinfo {author}
  {\bibfnamefont {Peide~D}\ \bibnamefont {Ye}},\ }\bibfield  {title} {\enquote
  {\bibinfo {title} {Phosphorene: an unexplored 2d semiconductor with a high
  hole mobility},}\ }\href@noop {} {\bibfield  {journal} {\bibinfo  {journal}
  {ACS Nano}\ }\textbf {\bibinfo {volume} {8}},\ \bibinfo {pages} {4033--4041}
  (\bibinfo {year} {2014})}\BibitemShut {NoStop}%
\bibitem [{\citenamefont {Castellanos-Gomez}\ \emph {et~al.}(2014)\citenamefont
  {Castellanos-Gomez}, \citenamefont {Vicarelli}, \citenamefont {Prada},
  \citenamefont {Island}, \citenamefont {Narasimha-Acharya}, \citenamefont
  {Blanter}, \citenamefont {Groenendijk}, \citenamefont {Buscema},
  \citenamefont {Steele}, \citenamefont {Alvarez}, \citenamefont {Zandbergen},
  \citenamefont {Palacios},\ and\ \citenamefont {van~der
  Zant}}]{Castellanos-Gomez:2M14}%
  \BibitemOpen
  \bibfield  {author} {\bibinfo {author} {\bibfnamefont {Andres}\ \bibnamefont
  {Castellanos-Gomez}}, \bibinfo {author} {\bibfnamefont {Leonardo}\
  \bibnamefont {Vicarelli}}, \bibinfo {author} {\bibfnamefont {Elsa}\
  \bibnamefont {Prada}}, \bibinfo {author} {\bibfnamefont {Joshua~O}\
  \bibnamefont {Island}}, \bibinfo {author} {\bibfnamefont {K~L}\ \bibnamefont
  {Narasimha-Acharya}}, \bibinfo {author} {\bibfnamefont {Sofya~I}\
  \bibnamefont {Blanter}}, \bibinfo {author} {\bibfnamefont {Dirk~J}\
  \bibnamefont {Groenendijk}}, \bibinfo {author} {\bibfnamefont {Michele}\
  \bibnamefont {Buscema}}, \bibinfo {author} {\bibfnamefont {Gary~A}\
  \bibnamefont {Steele}}, \bibinfo {author} {\bibfnamefont {J~V}\ \bibnamefont
  {Alvarez}}, \bibinfo {author} {\bibfnamefont {Henny~W}\ \bibnamefont
  {Zandbergen}}, \bibinfo {author} {\bibfnamefont {J~J}\ \bibnamefont
  {Palacios}}, \ and\ \bibinfo {author} {\bibfnamefont {Herre S~J}\
  \bibnamefont {van~der Zant}},\ }\bibfield  {title} {\enquote {\bibinfo
  {title} {Isolation and characterization of few-layer black phosphorus},}\
  }\href@noop {} {\bibfield  {journal} {\bibinfo  {journal} {2D Mater.}\
  }\textbf {\bibinfo {volume} {1}},\ \bibinfo {pages} {025001} (\bibinfo {year}
  {2014})}\BibitemShut {NoStop}%
\bibitem [{\citenamefont {Xia}\ \emph {et~al.}(2014{\natexlab{a}})\citenamefont
  {Xia}, \citenamefont {Wang},\ and\ \citenamefont {Jia}}]{Xia:NC14}%
  \BibitemOpen
  \bibfield  {author} {\bibinfo {author} {\bibfnamefont {Fengnian}\
  \bibnamefont {Xia}}, \bibinfo {author} {\bibfnamefont {Han}\ \bibnamefont
  {Wang}}, \ and\ \bibinfo {author} {\bibfnamefont {Yichen}\ \bibnamefont
  {Jia}},\ }\bibfield  {title} {\enquote {\bibinfo {title} {Rediscovering black
  phosphorus as an anisotropic layered material for optoelectronics and
  electronics},}\ }\href@noop {} {\bibfield  {journal} {\bibinfo  {journal}
  {Nature Comm.}\ }\textbf {\bibinfo {volume} {5}},\ \bibinfo {pages} {4458}
  (\bibinfo {year} {2014}{\natexlab{a}})}\BibitemShut {NoStop}%
\bibitem [{\citenamefont {Castellanos-Gomez}(2015)}]{Castellanos-Gomez:JPCL15}%
  \BibitemOpen
  \bibfield  {author} {\bibinfo {author} {\bibfnamefont {Andres}\ \bibnamefont
  {Castellanos-Gomez}},\ }\bibfield  {title} {\enquote {\bibinfo {title} {Black
  phosphorus: Narrow gap, wide applications},}\ }\href@noop {} {\bibfield
  {journal} {\bibinfo  {journal} {J. Phys. Chem. Lett.}\ }\textbf {\bibinfo
  {volume} {6}},\ \bibinfo {pages} {4280--4291} (\bibinfo {year}
  {2015})}\BibitemShut {NoStop}%
\bibitem [{\citenamefont {Guo}\ \emph {et~al.}(2016)\citenamefont {Guo},
  \citenamefont {Pospischil}, \citenamefont {Bhuiyan}, \citenamefont {Jiang},
  \citenamefont {Tian}, \citenamefont {Farmer}, \citenamefont {Deng},
  \citenamefont {Li}, \citenamefont {Han}, \citenamefont {Wang}, \citenamefont
  {Xia}, \citenamefont {Ma}, \citenamefont {Mueller},\ and\ \citenamefont
  {Xia}}]{Guo:16}%
  \BibitemOpen
  \bibfield  {author} {\bibinfo {author} {\bibfnamefont {Qiushi}\ \bibnamefont
  {Guo}}, \bibinfo {author} {\bibfnamefont {Andreas}\ \bibnamefont
  {Pospischil}}, \bibinfo {author} {\bibfnamefont {Maruf}\ \bibnamefont
  {Bhuiyan}}, \bibinfo {author} {\bibfnamefont {Hao}\ \bibnamefont {Jiang}},
  \bibinfo {author} {\bibfnamefont {He}~\bibnamefont {Tian}}, \bibinfo {author}
  {\bibfnamefont {Damon}\ \bibnamefont {Farmer}}, \bibinfo {author}
  {\bibfnamefont {Bingchen}\ \bibnamefont {Deng}}, \bibinfo {author}
  {\bibfnamefont {Cheng}\ \bibnamefont {Li}}, \bibinfo {author} {\bibfnamefont
  {Shu-Jen}\ \bibnamefont {Han}}, \bibinfo {author} {\bibfnamefont {Han}\
  \bibnamefont {Wang}}, \bibinfo {author} {\bibfnamefont {Qiangfei}\
  \bibnamefont {Xia}}, \bibinfo {author} {\bibfnamefont {Tso-Ping}\
  \bibnamefont {Ma}}, \bibinfo {author} {\bibfnamefont {Thomas}\ \bibnamefont
  {Mueller}}, \ and\ \bibinfo {author} {\bibfnamefont {Fengnian}\ \bibnamefont
  {Xia}},\ }\bibfield  {title} {\enquote {\bibinfo {title} {Black phosphorus
  mid-infrared photodetectors with high gain},}\ }\href@noop {} {\  (\bibinfo
  {year} {2016})},\ \Eprint {http://arxiv.org/abs/1603.07346}
  {arXiv:1603.07346} \BibitemShut {NoStop}%
\bibitem [{\citenamefont {Youngblood}\ \emph {et~al.}(2015)\citenamefont
  {Youngblood}, \citenamefont {Chen}, \citenamefont {Koester},\ and\
  \citenamefont {Li}}]{Youngblood:NP15}%
  \BibitemOpen
  \bibfield  {author} {\bibinfo {author} {\bibfnamefont {Nathan}\ \bibnamefont
  {Youngblood}}, \bibinfo {author} {\bibfnamefont {Che}\ \bibnamefont {Chen}},
  \bibinfo {author} {\bibfnamefont {Steven~J.}\ \bibnamefont {Koester}}, \ and\
  \bibinfo {author} {\bibfnamefont {Mo}~\bibnamefont {Li}},\ }\bibfield
  {title} {\enquote {\bibinfo {title} {Waveguide-integrated black phosphorus
  photodetector with high responsivity and low dark current},}\ }\href
  {http://dx.doi.org/10.1038/nphoton.2015.23} {\bibfield  {journal} {\bibinfo
  {journal} {Nat Photon}\ }\textbf {\bibinfo {volume} {9}},\ \bibinfo {pages}
  {247--252} (\bibinfo {year} {2015})}\BibitemShut {NoStop}%
\bibitem [{\citenamefont {Xia}\ \emph {et~al.}(2014{\natexlab{b}})\citenamefont
  {Xia}, \citenamefont {Wang}, \citenamefont {Xiao}, \citenamefont {Dubey},\
  and\ \citenamefont {Ramasubramaniam}}]{Xia:NP14}%
  \BibitemOpen
  \bibfield  {author} {\bibinfo {author} {\bibfnamefont {Fengnian}\
  \bibnamefont {Xia}}, \bibinfo {author} {\bibfnamefont {Han}\ \bibnamefont
  {Wang}}, \bibinfo {author} {\bibfnamefont {Di}~\bibnamefont {Xiao}}, \bibinfo
  {author} {\bibfnamefont {Madan}\ \bibnamefont {Dubey}}, \ and\ \bibinfo
  {author} {\bibfnamefont {Ashwin}\ \bibnamefont {Ramasubramaniam}},\
  }\bibfield  {title} {\enquote {\bibinfo {title} {Two-dimensional material
  nanophotonics},}\ }\href@noop {} {\bibfield  {journal} {\bibinfo  {journal}
  {Nature Photonics}\ }\textbf {\bibinfo {volume} {8}},\ \bibinfo {pages}
  {899--907} (\bibinfo {year} {2014}{\natexlab{b}})}\BibitemShut {NoStop}%
\bibitem [{\citenamefont {Ganesan}\ \emph {et~al.}(2016)\citenamefont
  {Ganesan}, \citenamefont {Linghu}, \citenamefont {Zhang}, \citenamefont
  {Feng},\ and\ \citenamefont {Shen}}]{Ganesan:APL16}%
  \BibitemOpen
  \bibfield  {author} {\bibinfo {author} {\bibfnamefont {Vellayappan
  Dheivanayagam~S/O}\ \bibnamefont {Ganesan}}, \bibinfo {author} {\bibfnamefont
  {Jiajun}\ \bibnamefont {Linghu}}, \bibinfo {author} {\bibfnamefont {Chun}\
  \bibnamefont {Zhang}}, \bibinfo {author} {\bibfnamefont {Yuan~Ping}\
  \bibnamefont {Feng}}, \ and\ \bibinfo {author} {\bibfnamefont {Lei}\
  \bibnamefont {Shen}},\ }\bibfield  {title} {\enquote {\bibinfo {title}
  {Heterostructures of phosphorene and transition metal dichalcogenides for
  excitonic solar cells: A first-principles study},}\ }\href@noop {} {\bibfield
   {journal} {\bibinfo  {journal} {Appl. Phys. Lett.}\ }\textbf {\bibinfo
  {volume} {108}},\ \bibinfo {pages} {122105} (\bibinfo {year}
  {2016})}\BibitemShut {NoStop}%
\bibitem [{\citenamefont {Castellanos-Gomez}(2016)}]{Castellanos-Gomez:NP16}%
  \BibitemOpen
  \bibfield  {author} {\bibinfo {author} {\bibfnamefont {Andres}\ \bibnamefont
  {Castellanos-Gomez}},\ }\bibfield  {title} {\enquote {\bibinfo {title} {Why
  all the fuss about 2d semiconductors?}}\ }\href@noop {} {\bibfield  {journal}
  {\bibinfo  {journal} {Nat Photon}\ }\textbf {\bibinfo {volume} {10}},\
  \bibinfo {pages} {202--204} (\bibinfo {year} {2016})}\BibitemShut {NoStop}%
\bibitem [{\citenamefont {Gillgren}\ \emph {et~al.}(2015)\citenamefont
  {Gillgren}, \citenamefont {Wickramaratne}, \citenamefont {Shi}, \citenamefont
  {Espiritu}, \citenamefont {Yang}, \citenamefont {Hu}, \citenamefont {Wei},
  \citenamefont {Liu}, \citenamefont {Mao}, \citenamefont {Watanabe},
  \citenamefont {Taniguchi}, \citenamefont {Bockrath}, \citenamefont {Barlas},
  \citenamefont {Lake},\ and\ \citenamefont {Lau}}]{Gillgren:2M15}%
  \BibitemOpen
  \bibfield  {author} {\bibinfo {author} {\bibfnamefont {Nathaniel}\
  \bibnamefont {Gillgren}}, \bibinfo {author} {\bibfnamefont {Darshana}\
  \bibnamefont {Wickramaratne}}, \bibinfo {author} {\bibfnamefont {Yanmeng}\
  \bibnamefont {Shi}}, \bibinfo {author} {\bibfnamefont {Tim}\ \bibnamefont
  {Espiritu}}, \bibinfo {author} {\bibfnamefont {Jiawei}\ \bibnamefont {Yang}},
  \bibinfo {author} {\bibfnamefont {Jin}\ \bibnamefont {Hu}}, \bibinfo {author}
  {\bibfnamefont {Jiang}\ \bibnamefont {Wei}}, \bibinfo {author} {\bibfnamefont
  {Xue}\ \bibnamefont {Liu}}, \bibinfo {author} {\bibfnamefont {Zhiqiang}\
  \bibnamefont {Mao}}, \bibinfo {author} {\bibfnamefont {Kenji}\ \bibnamefont
  {Watanabe}}, \bibinfo {author} {\bibfnamefont {Takashi}\ \bibnamefont
  {Taniguchi}}, \bibinfo {author} {\bibfnamefont {Marc}\ \bibnamefont
  {Bockrath}}, \bibinfo {author} {\bibfnamefont {Yafis}\ \bibnamefont
  {Barlas}}, \bibinfo {author} {\bibfnamefont {Roger~K}\ \bibnamefont {Lake}},
  \ and\ \bibinfo {author} {\bibfnamefont {Chun~Ning}\ \bibnamefont {Lau}},\
  }\bibfield  {title} {\enquote {\bibinfo {title} {Gate tunable quantum
  oscillations in air-stable and high mobility few-layer phosphorene
  heterostructures},}\ }\href@noop {} {\bibfield  {journal} {\bibinfo
  {journal} {2D Mater.}\ }\textbf {\bibinfo {volume} {2}},\ \bibinfo {pages}
  {011001} (\bibinfo {year} {2015})}\BibitemShut {NoStop}%
\bibitem [{\citenamefont {Chen}\ \emph {et~al.}(2015)\citenamefont {Chen},
  \citenamefont {Wu}, \citenamefont {Wu}, \citenamefont {Han}, \citenamefont
  {Xu}, \citenamefont {Wang}, \citenamefont {Ye}, \citenamefont {Han},
  \citenamefont {He}, \citenamefont {Cai},\ and\ \citenamefont
  {Wang}}]{Chen:NC15}%
  \BibitemOpen
  \bibfield  {author} {\bibinfo {author} {\bibfnamefont {Xiaolong}\
  \bibnamefont {Chen}}, \bibinfo {author} {\bibfnamefont {Yingying}\
  \bibnamefont {Wu}}, \bibinfo {author} {\bibfnamefont {Zefei}\ \bibnamefont
  {Wu}}, \bibinfo {author} {\bibfnamefont {Yu}~\bibnamefont {Han}}, \bibinfo
  {author} {\bibfnamefont {Shuigang}\ \bibnamefont {Xu}}, \bibinfo {author}
  {\bibfnamefont {Lin}\ \bibnamefont {Wang}}, \bibinfo {author} {\bibfnamefont
  {Weiguang}\ \bibnamefont {Ye}}, \bibinfo {author} {\bibfnamefont {Tianyi}\
  \bibnamefont {Han}}, \bibinfo {author} {\bibfnamefont {Yuheng}\ \bibnamefont
  {He}}, \bibinfo {author} {\bibfnamefont {Yuan}\ \bibnamefont {Cai}}, \ and\
  \bibinfo {author} {\bibfnamefont {Ning}\ \bibnamefont {Wang}},\ }\bibfield
  {title} {\enquote {\bibinfo {title} {High-quality sandwiched black phosphorus
  heterostructure and its quantum oscillations},}\ }\href@noop {} {\bibfield
  {journal} {\bibinfo  {journal} {Nat Commun}\ }\textbf {\bibinfo {volume}
  {6}},\ \bibinfo {pages} {7315} (\bibinfo {year} {2015})}\BibitemShut
  {NoStop}%
\bibitem [{\citenamefont {Li}\ \emph {et~al.}(2015{\natexlab{a}})\citenamefont
  {Li}, \citenamefont {Ye}, \citenamefont {Tran}, \citenamefont {Fei},
  \citenamefont {Chen}, \citenamefont {Wang}, \citenamefont {Wang},
  \citenamefont {Watanabe}, \citenamefont {Taniguchi}, \citenamefont {Yang},
  \citenamefont {Chen},\ and\ \citenamefont {Zhang}}]{Li:NN15}%
  \BibitemOpen
  \bibfield  {author} {\bibinfo {author} {\bibfnamefont {Likai}\ \bibnamefont
  {Li}}, \bibinfo {author} {\bibfnamefont {Guo~Jun}\ \bibnamefont {Ye}},
  \bibinfo {author} {\bibfnamefont {Vy}~\bibnamefont {Tran}}, \bibinfo {author}
  {\bibfnamefont {Ruixiang}\ \bibnamefont {Fei}}, \bibinfo {author}
  {\bibfnamefont {Guorui}\ \bibnamefont {Chen}}, \bibinfo {author}
  {\bibfnamefont {Huichao}\ \bibnamefont {Wang}}, \bibinfo {author}
  {\bibfnamefont {Jian}\ \bibnamefont {Wang}}, \bibinfo {author} {\bibfnamefont
  {Kenji}\ \bibnamefont {Watanabe}}, \bibinfo {author} {\bibfnamefont
  {Takashi}\ \bibnamefont {Taniguchi}}, \bibinfo {author} {\bibfnamefont
  {Li}~\bibnamefont {Yang}}, \bibinfo {author} {\bibfnamefont {Xian~Hui}\
  \bibnamefont {Chen}}, \ and\ \bibinfo {author} {\bibfnamefont {Yuanbo}\
  \bibnamefont {Zhang}},\ }\bibfield  {title} {\enquote {\bibinfo {title}
  {Quantum oscillations in a two-dimensional electron gas in black phosphorus
  thin films},}\ }\href@noop {} {\bibfield  {journal} {\bibinfo  {journal} {Nat
  Nano}\ }\textbf {\bibinfo {volume} {10}},\ \bibinfo {pages} {608--613}
  (\bibinfo {year} {2015}{\natexlab{a}})}\BibitemShut {NoStop}%
\bibitem [{\citenamefont {Rold\'an}\ \emph {et~al.}(2015)\citenamefont
  {Rold\'an}, \citenamefont {Castellanos-Gomez}, \citenamefont {Cappelluti},\
  and\ \citenamefont {Guinea}}]{Roldan:JPCM15}%
  \BibitemOpen
  \bibfield  {author} {\bibinfo {author} {\bibfnamefont {Rafael}\ \bibnamefont
  {Rold\'an}}, \bibinfo {author} {\bibfnamefont {Andr\'es}\ \bibnamefont
  {Castellanos-Gomez}}, \bibinfo {author} {\bibfnamefont {Emmanuele}\
  \bibnamefont {Cappelluti}}, \ and\ \bibinfo {author} {\bibfnamefont
  {Francisco}\ \bibnamefont {Guinea}},\ }\bibfield  {title} {\enquote {\bibinfo
  {title} {Strain engineering in semiconducting two-dimensional crystals},}\
  }\href@noop {} {\bibfield  {journal} {\bibinfo  {journal} {J. Physics:
  Condens. Matter}\ }\textbf {\bibinfo {volume} {27}},\ \bibinfo {pages}
  {313201} (\bibinfo {year} {2015})}\BibitemShut {NoStop}%
\bibitem [{\citenamefont {Chang}\ \emph {et~al.}(2013)\citenamefont {Chang},
  \citenamefont {Fan}, \citenamefont {Lin},\ and\ \citenamefont
  {Kuo}}]{Chang:PRB13}%
  \BibitemOpen
  \bibfield  {author} {\bibinfo {author} {\bibfnamefont {Chung-Huai}\
  \bibnamefont {Chang}}, \bibinfo {author} {\bibfnamefont {Xiaofeng}\
  \bibnamefont {Fan}}, \bibinfo {author} {\bibfnamefont {Shi-Hsin}\
  \bibnamefont {Lin}}, \ and\ \bibinfo {author} {\bibfnamefont {Jer-Lai}\
  \bibnamefont {Kuo}},\ }\bibfield  {title} {\enquote {\bibinfo {title}
  {Orbital analysis of electronic structure and phonon dispersion in
  mos${}_{2}$, mose${}_{2}$, ws${}_{2}$, and wse${}_{2}$ monolayers under
  strain},}\ }\href@noop {} {\bibfield  {journal} {\bibinfo  {journal} {Phys.
  Rev. B}\ }\textbf {\bibinfo {volume} {88}},\ \bibinfo {pages} {195420}
  (\bibinfo {year} {2013})}\BibitemShut {NoStop}%
\bibitem [{\citenamefont {Peng}\ \emph {et~al.}(2014)\citenamefont {Peng},
  \citenamefont {Wei},\ and\ \citenamefont {Copple}}]{Peng:PRB14}%
  \BibitemOpen
  \bibfield  {author} {\bibinfo {author} {\bibfnamefont {Xihong}\ \bibnamefont
  {Peng}}, \bibinfo {author} {\bibfnamefont {Qun}\ \bibnamefont {Wei}}, \ and\
  \bibinfo {author} {\bibfnamefont {Andrew}\ \bibnamefont {Copple}},\
  }\bibfield  {title} {\enquote {\bibinfo {title} {Strain-engineered
  direct-indirect band gap transition and its mechanism in two-dimensional
  phosphorene},}\ }\href@noop {} {\bibfield  {journal} {\bibinfo  {journal}
  {Phys. Rev. B}\ }\textbf {\bibinfo {volume} {90}},\ \bibinfo {pages} {085402}
  (\bibinfo {year} {2014})}\BibitemShut {NoStop}%
\bibitem [{\citenamefont {Quereda}\ \emph {et~al.}(2016)\citenamefont
  {Quereda}, \citenamefont {San-Jose}, \citenamefont {Parente}, \citenamefont
  {Vaquero-Garzon}, \citenamefont {Molina-Mendoza}, \citenamefont
  {Agra{\"\i}t}, \citenamefont {Rubio-Bollinger}, \citenamefont {Guinea},
  \citenamefont {Rold{\'a}n},\ and\ \citenamefont
  {Castellanos-Gomez}}]{Quereda:NL16}%
  \BibitemOpen
  \bibfield  {author} {\bibinfo {author} {\bibfnamefont {Jorge}\ \bibnamefont
  {Quereda}}, \bibinfo {author} {\bibfnamefont {Pablo}\ \bibnamefont
  {San-Jose}}, \bibinfo {author} {\bibfnamefont {Vincenzo}\ \bibnamefont
  {Parente}}, \bibinfo {author} {\bibfnamefont {Luis}\ \bibnamefont
  {Vaquero-Garzon}}, \bibinfo {author} {\bibfnamefont {Aday~J.}\ \bibnamefont
  {Molina-Mendoza}}, \bibinfo {author} {\bibfnamefont {Nicol{\'a}s}\
  \bibnamefont {Agra{\"\i}t}}, \bibinfo {author} {\bibfnamefont {Gabino}\
  \bibnamefont {Rubio-Bollinger}}, \bibinfo {author} {\bibfnamefont
  {Francisco}\ \bibnamefont {Guinea}}, \bibinfo {author} {\bibfnamefont
  {Rafael}\ \bibnamefont {Rold{\'a}n}}, \ and\ \bibinfo {author} {\bibfnamefont
  {Andres}\ \bibnamefont {Castellanos-Gomez}},\ }\bibfield  {title} {\enquote
  {\bibinfo {title} {Strong modulation of optical properties in black
  phosphorus through strain-engineered rippling},}\ }\href
  {http://dx.doi.org/10.1021/acs.nanolett.5b04670} {\bibfield  {journal}
  {\bibinfo  {journal} {Nano Letters}\ }\textbf {\bibinfo {volume} {16}},\
  \bibinfo {pages} {2931--2937} (\bibinfo {year} {2016})}\BibitemShut {NoStop}%
\bibitem [{\citenamefont {Fei}\ and\ \citenamefont {Yang}(2014)}]{Fei:NL14}%
  \BibitemOpen
  \bibfield  {author} {\bibinfo {author} {\bibfnamefont {Ruixiang}\
  \bibnamefont {Fei}}\ and\ \bibinfo {author} {\bibfnamefont {Li}~\bibnamefont
  {Yang}},\ }\bibfield  {title} {\enquote {\bibinfo {title} {Strain-engineering
  the anisotropic electrical conductance of few-layer black phosphorus},}\
  }\href@noop {} {\bibfield  {journal} {\bibinfo  {journal} {Nano Lett.}\
  }\textbf {\bibinfo {volume} {14}},\ \bibinfo {pages} {2884--2889} (\bibinfo
  {year} {2014})}\BibitemShut {NoStop}%
\bibitem [{\citenamefont {Moody}\ \emph {et~al.}(2016)\citenamefont {Moody},
  \citenamefont {Schaibley},\ and\ \citenamefont {Xu}}]{Moody:16}%
  \BibitemOpen
  \bibfield  {author} {\bibinfo {author} {\bibfnamefont {Galan}\ \bibnamefont
  {Moody}}, \bibinfo {author} {\bibfnamefont {John}\ \bibnamefont {Schaibley}},
  \ and\ \bibinfo {author} {\bibfnamefont {Xiaodong}\ \bibnamefont {Xu}},\
  }\bibfield  {title} {\enquote {\bibinfo {title} {Exciton dynamics in
  monolayer transition metal dichalcogenides},}\ }\href@noop {} {\  (\bibinfo
  {year} {2016})},\ \Eprint {http://arxiv.org/abs/1602.04543}
  {arXiv:1602.04543} \BibitemShut {NoStop}%
\bibitem [{\citenamefont {Splendiani}\ \emph {et~al.}(2010)\citenamefont
  {Splendiani}, \citenamefont {Sun}, \citenamefont {Zhang}, \citenamefont {Li},
  \citenamefont {Kim}, \citenamefont {Chim}, \citenamefont {Galli},\ and\
  \citenamefont {Wang}}]{Splendiani:NL10}%
  \BibitemOpen
  \bibfield  {author} {\bibinfo {author} {\bibfnamefont {Andrea}\ \bibnamefont
  {Splendiani}}, \bibinfo {author} {\bibfnamefont {Liang}\ \bibnamefont {Sun}},
  \bibinfo {author} {\bibfnamefont {Yuanbo}\ \bibnamefont {Zhang}}, \bibinfo
  {author} {\bibfnamefont {Tianshu}\ \bibnamefont {Li}}, \bibinfo {author}
  {\bibfnamefont {Jonghwan}\ \bibnamefont {Kim}}, \bibinfo {author}
  {\bibfnamefont {Chi-Yung}\ \bibnamefont {Chim}}, \bibinfo {author}
  {\bibfnamefont {Giulia}\ \bibnamefont {Galli}}, \ and\ \bibinfo {author}
  {\bibfnamefont {Feng}\ \bibnamefont {Wang}},\ }\bibfield  {title} {\enquote
  {\bibinfo {title} {Emerging photoluminescence in monolayer mos2},}\
  }\href@noop {} {\bibfield  {journal} {\bibinfo  {journal} {Nano Lett.}\
  }\textbf {\bibinfo {volume} {10}},\ \bibinfo {pages} {1271--1275} (\bibinfo
  {year} {2010})}\BibitemShut {NoStop}%
\bibitem [{\citenamefont {Mak}\ \emph {et~al.}(2012)\citenamefont {Mak},
  \citenamefont {He}, \citenamefont {Shan},\ and\ \citenamefont
  {Heinz}}]{Mak:NN12}%
  \BibitemOpen
  \bibfield  {author} {\bibinfo {author} {\bibfnamefont {Kin~Fai}\ \bibnamefont
  {Mak}}, \bibinfo {author} {\bibfnamefont {Keliang}\ \bibnamefont {He}},
  \bibinfo {author} {\bibfnamefont {Jie}\ \bibnamefont {Shan}}, \ and\ \bibinfo
  {author} {\bibfnamefont {Tony~F}\ \bibnamefont {Heinz}},\ }\bibfield  {title}
  {\enquote {\bibinfo {title} {Control of valley polarization in monolayer mos2
  by optical helicity},}\ }\href@noop {} {\bibfield  {journal} {\bibinfo
  {journal} {Nature Nano.}\ }\textbf {\bibinfo {volume} {7}},\ \bibinfo {pages}
  {494--498} (\bibinfo {year} {2012})}\BibitemShut {NoStop}%
\bibitem [{\citenamefont {Ugeda}\ \emph {et~al.}(2014)\citenamefont {Ugeda},
  \citenamefont {Bradley}, \citenamefont {Shi}, \citenamefont {da~Jornada},
  \citenamefont {Zhang}, \citenamefont {Qiu}, \citenamefont {Ruan},
  \citenamefont {Mo}, \citenamefont {Hussain}, \citenamefont {Shen},
  \citenamefont {Wang}, \citenamefont {Louie},\ and\ \citenamefont
  {Crommie}}]{Ugeda:NM14}%
  \BibitemOpen
  \bibfield  {author} {\bibinfo {author} {\bibfnamefont {Miguel~M.}\
  \bibnamefont {Ugeda}}, \bibinfo {author} {\bibfnamefont {Aaron~J.}\
  \bibnamefont {Bradley}}, \bibinfo {author} {\bibfnamefont {Su-Fei}\
  \bibnamefont {Shi}}, \bibinfo {author} {\bibfnamefont {Felipe~H.}\
  \bibnamefont {da~Jornada}}, \bibinfo {author} {\bibfnamefont
  {Yi}~\bibnamefont {Zhang}}, \bibinfo {author} {\bibfnamefont {Diana~Y.}\
  \bibnamefont {Qiu}}, \bibinfo {author} {\bibfnamefont {Wei}\ \bibnamefont
  {Ruan}}, \bibinfo {author} {\bibfnamefont {Sung-Kwan}\ \bibnamefont {Mo}},
  \bibinfo {author} {\bibfnamefont {Zahid}\ \bibnamefont {Hussain}}, \bibinfo
  {author} {\bibfnamefont {Zhi-Xun}\ \bibnamefont {Shen}}, \bibinfo {author}
  {\bibfnamefont {Feng}\ \bibnamefont {Wang}}, \bibinfo {author} {\bibfnamefont
  {Steven~G.}\ \bibnamefont {Louie}}, \ and\ \bibinfo {author} {\bibfnamefont
  {Michael~F.}\ \bibnamefont {Crommie}},\ }\bibfield  {title} {\enquote
  {\bibinfo {title} {Giant bandgap renormalization and excitonic effects in a
  monolayer transition metal dichalcogenide semiconductor},}\ }\href@noop {}
  {\bibfield  {journal} {\bibinfo  {journal} {Nat Mater}\ }\textbf {\bibinfo
  {volume} {13}},\ \bibinfo {pages} {1091--1095} (\bibinfo {year}
  {2014})}\BibitemShut {NoStop}%
\bibitem [{\citenamefont {Wang}\ \emph {et~al.}(2015)\citenamefont {Wang},
  \citenamefont {Jones}, \citenamefont {Seyler}, \citenamefont {Tran},
  \citenamefont {Jia}, \citenamefont {Zhao}, \citenamefont {Wang},
  \citenamefont {Yang}, \citenamefont {Xu},\ and\ \citenamefont
  {Xia}}]{Wang:NN15}%
  \BibitemOpen
  \bibfield  {author} {\bibinfo {author} {\bibfnamefont {Xiaomu}\ \bibnamefont
  {Wang}}, \bibinfo {author} {\bibfnamefont {Aaron~M.}\ \bibnamefont {Jones}},
  \bibinfo {author} {\bibfnamefont {Kyle~L.}\ \bibnamefont {Seyler}}, \bibinfo
  {author} {\bibfnamefont {Vy}~\bibnamefont {Tran}}, \bibinfo {author}
  {\bibfnamefont {Yichen}\ \bibnamefont {Jia}}, \bibinfo {author}
  {\bibfnamefont {Huan}\ \bibnamefont {Zhao}}, \bibinfo {author} {\bibfnamefont
  {Han}\ \bibnamefont {Wang}}, \bibinfo {author} {\bibfnamefont
  {Li}~\bibnamefont {Yang}}, \bibinfo {author} {\bibfnamefont {Xiaodong}\
  \bibnamefont {Xu}}, \ and\ \bibinfo {author} {\bibfnamefont {Fengnian}\
  \bibnamefont {Xia}},\ }\bibfield  {title} {\enquote {\bibinfo {title} {Highly
  anisotropic and robust excitons in monolayer black phosphorus},}\ }\href@noop
  {} {\bibfield  {journal} {\bibinfo  {journal} {Nat Nano}\ }\textbf {\bibinfo
  {volume} {10}},\ \bibinfo {pages} {517--521} (\bibinfo {year}
  {2015})}\BibitemShut {NoStop}%
\bibitem [{\citenamefont {Yang}\ \emph {et~al.}(2015)\citenamefont {Yang},
  \citenamefont {Xu}, \citenamefont {Pei}, \citenamefont {Myint}, \citenamefont
  {Wang}, \citenamefont {Wang}, \citenamefont {Zhang}, \citenamefont {Yu},\
  and\ \citenamefont {Lu}}]{Yang:LSA15}%
  \BibitemOpen
  \bibfield  {author} {\bibinfo {author} {\bibfnamefont {Jiong}\ \bibnamefont
  {Yang}}, \bibinfo {author} {\bibfnamefont {Renjing}\ \bibnamefont {Xu}},
  \bibinfo {author} {\bibfnamefont {Jiajie}\ \bibnamefont {Pei}}, \bibinfo
  {author} {\bibfnamefont {Ye~Win}\ \bibnamefont {Myint}}, \bibinfo {author}
  {\bibfnamefont {Fan}\ \bibnamefont {Wang}}, \bibinfo {author} {\bibfnamefont
  {Zhu}\ \bibnamefont {Wang}}, \bibinfo {author} {\bibfnamefont {Shuang}\
  \bibnamefont {Zhang}}, \bibinfo {author} {\bibfnamefont {Zongfu}\
  \bibnamefont {Yu}}, \ and\ \bibinfo {author} {\bibfnamefont {Yuerui}\
  \bibnamefont {Lu}},\ }\bibfield  {title} {\enquote {\bibinfo {title} {Optical
  tuning of exciton and trion emissions in monolayer phosphorene},}\
  }\href@noop {} {\bibfield  {journal} {\bibinfo  {journal} {Light Sci Appl}\
  }\textbf {\bibinfo {volume} {4}},\ \bibinfo {pages} {e312--} (\bibinfo {year}
  {2015})}\BibitemShut {NoStop}%
\bibitem [{Note1()}]{Note1}%
  \BibitemOpen
  \bibinfo {note} {{Note that in gapless graphene, excited electrons and holes
  give rise to a photocurrent through the thermoelectric effect \cite
  {Gabor648}. In contrast, the funnelling of (neutral) excitons considered here
  is driven by the potential gradient associated to the changing gap, and it
  does not involve charge currents.}}\BibitemShut {Stop}%
\bibitem [{\citenamefont {Shockley}\ and\ \citenamefont
  {Queisser}(1961)}]{Shockley:JAP61}%
  \BibitemOpen
  \bibfield  {author} {\bibinfo {author} {\bibfnamefont {William}\ \bibnamefont
  {Shockley}}\ and\ \bibinfo {author} {\bibfnamefont {Hans~J.}\ \bibnamefont
  {Queisser}},\ }\bibfield  {title} {\enquote {\bibinfo {title} {Detailed
  balance limit of efficiency of p?n junction solar cells},}\ }\href@noop {}
  {\bibfield  {journal} {\bibinfo  {journal} {J. Appl. Phys.}\ }\textbf
  {\bibinfo {volume} {32}},\ \bibinfo {pages} {510--519} (\bibinfo {year}
  {1961})}\BibitemShut {NoStop}%
\bibitem [{\citenamefont {Castellanos-Gomez}\ \emph {et~al.}(2013)\citenamefont
  {Castellanos-Gomez}, \citenamefont {Rold{\'a}n}, \citenamefont {Cappelluti},
  \citenamefont {Buscema}, \citenamefont {Guinea}, \citenamefont {van~der
  Zant},\ and\ \citenamefont {Steele}}]{Castellanos-Gomez:NL13}%
  \BibitemOpen
  \bibfield  {author} {\bibinfo {author} {\bibfnamefont {Andres}\ \bibnamefont
  {Castellanos-Gomez}}, \bibinfo {author} {\bibfnamefont {Rafael}\ \bibnamefont
  {Rold{\'a}n}}, \bibinfo {author} {\bibfnamefont {Emmanuele}\ \bibnamefont
  {Cappelluti}}, \bibinfo {author} {\bibfnamefont {Michele}\ \bibnamefont
  {Buscema}}, \bibinfo {author} {\bibfnamefont {Francisco}\ \bibnamefont
  {Guinea}}, \bibinfo {author} {\bibfnamefont {Herre S.~J.}\ \bibnamefont
  {van~der Zant}}, \ and\ \bibinfo {author} {\bibfnamefont {Gary~A.}\
  \bibnamefont {Steele}},\ }\bibfield  {title} {\enquote {\bibinfo {title}
  {Local strain engineering in atomically thin mos2},}\ }\href@noop {}
  {\bibfield  {journal} {\bibinfo  {journal} {Nano Lett.}\ }\textbf {\bibinfo
  {volume} {13}},\ \bibinfo {pages} {5361--5366} (\bibinfo {year}
  {2013})}\BibitemShut {NoStop}%
\bibitem [{\citenamefont {Li}\ \emph {et~al.}(2015{\natexlab{b}})\citenamefont
  {Li}, \citenamefont {Contryman}, \citenamefont {Qian}, \citenamefont
  {Ardakani}, \citenamefont {Gong}, \citenamefont {Wang}, \citenamefont
  {Weisse}, \citenamefont {Lee}, \citenamefont {Zhao}, \citenamefont {Ajayan},
  \citenamefont {Li}, \citenamefont {Manoharan},\ and\ \citenamefont
  {Zheng}}]{Li:NC15}%
  \BibitemOpen
  \bibfield  {author} {\bibinfo {author} {\bibfnamefont {Hong}\ \bibnamefont
  {Li}}, \bibinfo {author} {\bibfnamefont {Alex~W.}\ \bibnamefont {Contryman}},
  \bibinfo {author} {\bibfnamefont {Xiaofeng}\ \bibnamefont {Qian}}, \bibinfo
  {author} {\bibfnamefont {Sina~Moeini}\ \bibnamefont {Ardakani}}, \bibinfo
  {author} {\bibfnamefont {Yongji}\ \bibnamefont {Gong}}, \bibinfo {author}
  {\bibfnamefont {Xingli}\ \bibnamefont {Wang}}, \bibinfo {author}
  {\bibfnamefont {Jeffrey~M.}\ \bibnamefont {Weisse}}, \bibinfo {author}
  {\bibfnamefont {Chi~Hwan}\ \bibnamefont {Lee}}, \bibinfo {author}
  {\bibfnamefont {Jiheng}\ \bibnamefont {Zhao}}, \bibinfo {author}
  {\bibfnamefont {Pulickel~M.}\ \bibnamefont {Ajayan}}, \bibinfo {author}
  {\bibfnamefont {Ju}~\bibnamefont {Li}}, \bibinfo {author} {\bibfnamefont
  {Hari~C.}\ \bibnamefont {Manoharan}}, \ and\ \bibinfo {author} {\bibfnamefont
  {Xiaolin}\ \bibnamefont {Zheng}},\ }\bibfield  {title} {\enquote {\bibinfo
  {title} {Optoelectronic crystal of artificial atoms in strain-textured
  molybdenum disulphide},}\ }\href@noop {} {\bibfield  {journal} {\bibinfo
  {journal} {Nat. Commun.}\ }\textbf {\bibinfo {volume} {6}},\ \bibinfo {pages}
  {7381} (\bibinfo {year} {2015}{\natexlab{b}})}\BibitemShut {NoStop}%
\bibitem [{\citenamefont {Jariwala}\ \emph {et~al.}(2014)\citenamefont
  {Jariwala}, \citenamefont {Sangwan}, \citenamefont {Lauhon}, \citenamefont
  {Marks},\ and\ \citenamefont {Hersam}}]{Jariwala:AN14}%
  \BibitemOpen
  \bibfield  {author} {\bibinfo {author} {\bibfnamefont {Deep}\ \bibnamefont
  {Jariwala}}, \bibinfo {author} {\bibfnamefont {Vinod~K.}\ \bibnamefont
  {Sangwan}}, \bibinfo {author} {\bibfnamefont {Lincoln~J.}\ \bibnamefont
  {Lauhon}}, \bibinfo {author} {\bibfnamefont {Tobin~J.}\ \bibnamefont
  {Marks}}, \ and\ \bibinfo {author} {\bibfnamefont {Mark~C.}\ \bibnamefont
  {Hersam}},\ }\bibfield  {title} {\enquote {\bibinfo {title} {Emerging device
  applications for semiconducting two-dimensional transition metal
  dichalcogenides},}\ }\href@noop {} {\bibfield  {journal} {\bibinfo  {journal}
  {ACS Nano}\ }\textbf {\bibinfo {volume} {8}},\ \bibinfo {pages} {1102--1120}
  (\bibinfo {year} {2014})}\BibitemShut {NoStop}%
\bibitem [{\citenamefont {Yu}\ \emph {et~al.}(2014)\citenamefont {Yu},
  \citenamefont {Feng},\ and\ \citenamefont {Hone}}]{Yu:MB14}%
  \BibitemOpen
  \bibfield  {author} {\bibinfo {author} {\bibfnamefont {Dapeng}\ \bibnamefont
  {Yu}}, \bibinfo {author} {\bibfnamefont {Ji}~\bibnamefont {Feng}}, \ and\
  \bibinfo {author} {\bibfnamefont {James}\ \bibnamefont {Hone}},\ }\bibfield
  {title} {\enquote {\bibinfo {title} {Elastically strained nanowires and
  atomic sheets},}\ }\href@noop {} {\bibfield  {journal} {\bibinfo  {journal}
  {MRS Bull.}\ }\textbf {\bibinfo {volume} {39}},\ \bibinfo {pages} {157--162}
  (\bibinfo {year} {2014})}\BibitemShut {NoStop}%
\bibitem [{\citenamefont {Li}\ \emph {et~al.}(2014{\natexlab{b}})\citenamefont
  {Li}, \citenamefont {Shan},\ and\ \citenamefont {Ma}}]{Li:MB14}%
  \BibitemOpen
  \bibfield  {author} {\bibinfo {author} {\bibfnamefont {Ju}~\bibnamefont
  {Li}}, \bibinfo {author} {\bibfnamefont {Zhiwei}\ \bibnamefont {Shan}}, \
  and\ \bibinfo {author} {\bibfnamefont {Evan}\ \bibnamefont {Ma}},\ }\bibfield
   {title} {\enquote {\bibinfo {title} {Elastic strain engineering for
  unprecedented materials properties},}\ }\href@noop {} {\bibfield  {journal}
  {\bibinfo  {journal} {MRS Bull.}\ }\textbf {\bibinfo {volume} {39}},\
  \bibinfo {pages} {108--114} (\bibinfo {year}
  {2014}{\natexlab{b}})}\BibitemShut {NoStop}%
\bibitem [{\citenamefont {Rudenko}\ \emph {et~al.}(2015)\citenamefont
  {Rudenko}, \citenamefont {Yuan},\ and\ \citenamefont
  {Katsnelson}}]{Rudenko:PRB15}%
  \BibitemOpen
  \bibfield  {author} {\bibinfo {author} {\bibfnamefont {A.~N.}\ \bibnamefont
  {Rudenko}}, \bibinfo {author} {\bibfnamefont {Shengjun}\ \bibnamefont
  {Yuan}}, \ and\ \bibinfo {author} {\bibfnamefont {M.~I.}\ \bibnamefont
  {Katsnelson}},\ }\bibfield  {title} {\enquote {\bibinfo {title} {Toward a
  realistic description of multilayer black phosphorus: From $gw$ approximation
  to large-scale tight-binding simulations},}\ }\href@noop {} {\bibfield
  {journal} {\bibinfo  {journal} {Phys. Rev. B}\ }\textbf {\bibinfo {volume}
  {92}},\ \bibinfo {pages} {085419} (\bibinfo {year} {2015})}\BibitemShut
  {NoStop}%
\bibitem [{\citenamefont {Cappelluti}\ \emph {et~al.}(2013)\citenamefont
  {Cappelluti}, \citenamefont {Rold\'an}, \citenamefont {Silva-Guill\'en},
  \citenamefont {Ordej\'on},\ and\ \citenamefont {Guinea}}]{Cappelluti:PRB13}%
  \BibitemOpen
  \bibfield  {author} {\bibinfo {author} {\bibfnamefont {E.}~\bibnamefont
  {Cappelluti}}, \bibinfo {author} {\bibfnamefont {R.}~\bibnamefont
  {Rold\'an}}, \bibinfo {author} {\bibfnamefont {J.~A.}\ \bibnamefont
  {Silva-Guill\'en}}, \bibinfo {author} {\bibfnamefont {P.}~\bibnamefont
  {Ordej\'on}}, \ and\ \bibinfo {author} {\bibfnamefont {F.}~\bibnamefont
  {Guinea}},\ }\bibfield  {title} {\enquote {\bibinfo {title} {Tight-binding
  model and direct-gap/indirect-gap transition in single-layer and multilayer
  mos${}_{2}$},}\ }\href {\doibase 10.1103/PhysRevB.88.075409} {\bibfield
  {journal} {\bibinfo  {journal} {Phys. Rev. B}\ }\textbf {\bibinfo {volume}
  {88}},\ \bibinfo {pages} {075409} (\bibinfo {year} {2013})}\BibitemShut
  {NoStop}%
\bibitem [{Note2()}]{Note2}%
  \BibitemOpen
  \bibinfo {note} {The gap for MoS$_2$ is underestimated by about $20-25\%$ in
  our model as compared to most experiments. This is expected, as the
  tight-binding for MoS$_2$, taken from Ref. \cite {Cappelluti:PRB13}, was
  fitted to LDA calculations, which are known to underestimate gaps \cite
  {Perdew:IJQC85,Qiu:PRL13}, as opposed to the GW-LDA used for BP \cite
  {Rudenko:PRB15}. This in turn is expected to lead to an overestimation of
  exciton lifetimes and funnel drift lengths in MoS$_2$.}\BibitemShut {Stop}%
\bibitem [{\citenamefont {Rodin}\ \emph {et~al.}(2014)\citenamefont {Rodin},
  \citenamefont {Carvalho},\ and\ \citenamefont {Castro~Neto}}]{Rodin:PRB14}%
  \BibitemOpen
  \bibfield  {author} {\bibinfo {author} {\bibfnamefont {A.~S.}\ \bibnamefont
  {Rodin}}, \bibinfo {author} {\bibfnamefont {A.}~\bibnamefont {Carvalho}}, \
  and\ \bibinfo {author} {\bibfnamefont {A.~H.}\ \bibnamefont {Castro~Neto}},\
  }\bibfield  {title} {\enquote {\bibinfo {title} {Excitons in anisotropic
  two-dimensional semiconducting crystals},}\ }\href@noop {} {\bibfield
  {journal} {\bibinfo  {journal} {Phys. Rev. B}\ }\textbf {\bibinfo {volume}
  {90}},\ \bibinfo {pages} {075429} (\bibinfo {year} {2014})}\BibitemShut
  {NoStop}%
\bibitem [{\citenamefont {Tran}\ \emph {et~al.}(2014)\citenamefont {Tran},
  \citenamefont {Soklaski}, \citenamefont {Liang},\ and\ \citenamefont
  {Yang}}]{Tran:PRB14}%
  \BibitemOpen
  \bibfield  {author} {\bibinfo {author} {\bibfnamefont {Vy}~\bibnamefont
  {Tran}}, \bibinfo {author} {\bibfnamefont {Ryan}\ \bibnamefont {Soklaski}},
  \bibinfo {author} {\bibfnamefont {Yufeng}\ \bibnamefont {Liang}}, \ and\
  \bibinfo {author} {\bibfnamefont {Li}~\bibnamefont {Yang}},\ }\bibfield
  {title} {\enquote {\bibinfo {title} {Layer-controlled band gap and
  anisotropic excitons in few-layer black phosphorus},}\ }\href@noop {}
  {\bibfield  {journal} {\bibinfo  {journal} {Phys. Rev. B}\ }\textbf {\bibinfo
  {volume} {89}},\ \bibinfo {pages} {235319} (\bibinfo {year}
  {2014})}\BibitemShut {NoStop}%
\bibitem [{\citenamefont {Prada}\ \emph {et~al.}(2015)\citenamefont {Prada},
  \citenamefont {Alvarez}, \citenamefont {Narasimha-Acharya}, \citenamefont
  {Bailen},\ and\ \citenamefont {Palacios}}]{Prada:PRB15}%
  \BibitemOpen
  \bibfield  {author} {\bibinfo {author} {\bibfnamefont {Elsa}\ \bibnamefont
  {Prada}}, \bibinfo {author} {\bibfnamefont {J.~V.}\ \bibnamefont {Alvarez}},
  \bibinfo {author} {\bibfnamefont {K.~L.}\ \bibnamefont {Narasimha-Acharya}},
  \bibinfo {author} {\bibfnamefont {F.~J.}\ \bibnamefont {Bailen}}, \ and\
  \bibinfo {author} {\bibfnamefont {J.~J.}\ \bibnamefont {Palacios}},\
  }\bibfield  {title} {\enquote {\bibinfo {title} {Effective-mass theory for
  the anisotropic exciton in two-dimensional crystals: Application to
  phosphorene},}\ }\href@noop {} {\bibfield  {journal} {\bibinfo  {journal}
  {Phys. Rev. B}\ }\textbf {\bibinfo {volume} {91}},\ \bibinfo {pages} {245421}
  (\bibinfo {year} {2015})}\BibitemShut {NoStop}%
\bibitem [{\citenamefont {Chaves}\ \emph {et~al.}(2015)\citenamefont {Chaves},
  \citenamefont {Low}, \citenamefont {Avouris}, \citenamefont {{\c C}ak{\i}r},\
  and\ \citenamefont {Peeters}}]{Chaves:PRB15}%
  \BibitemOpen
  \bibfield  {author} {\bibinfo {author} {\bibfnamefont {A.}~\bibnamefont
  {Chaves}}, \bibinfo {author} {\bibfnamefont {Tony}\ \bibnamefont {Low}},
  \bibinfo {author} {\bibfnamefont {P.}~\bibnamefont {Avouris}}, \bibinfo
  {author} {\bibfnamefont {D.}~\bibnamefont {{\c C}ak{\i}r}}, \ and\ \bibinfo
  {author} {\bibfnamefont {F.~M.}\ \bibnamefont {Peeters}},\ }\bibfield
  {title} {\enquote {\bibinfo {title} {Anisotropic exciton stark shift in black
  phosphorus},}\ }\href@noop {} {\bibfield  {journal} {\bibinfo  {journal}
  {Phys. Rev. B}\ }\textbf {\bibinfo {volume} {91}},\ \bibinfo {pages} {155311}
  (\bibinfo {year} {2015})}\BibitemShut {NoStop}%
\bibitem [{\citenamefont {Yu}\ and\ \citenamefont {Cardona}(2005)}]{Yu:05}%
  \BibitemOpen
  \bibfield  {author} {\bibinfo {author} {\bibfnamefont {Peter~Y}\ \bibnamefont
  {Yu}}\ and\ \bibinfo {author} {\bibfnamefont {Manuel}\ \bibnamefont
  {Cardona}},\ }\href@noop {} {\emph {\bibinfo {title} {Fundamentals of
  semiconductors}}}\ (\bibinfo  {publisher} {Springer},\ \bibinfo {year}
  {2005})\BibitemShut {NoStop}%
\bibitem [{\citenamefont {Keldysh}(1979)}]{Keldysh:JL79}%
  \BibitemOpen
  \bibfield  {author} {\bibinfo {author} {\bibfnamefont {L.~V.}\ \bibnamefont
  {Keldysh}},\ }\bibfield  {title} {\enquote {\bibinfo {title} {Coulomb
  interaction in thin semiconductors and semimetals films},}\ }\href@noop {}
  {\bibfield  {journal} {\bibinfo  {journal} {JETP Lett.}\ }\textbf {\bibinfo
  {volume} {29}},\ \bibinfo {pages} {658} (\bibinfo {year} {1979})}\BibitemShut
  {NoStop}%
\bibitem [{\citenamefont {Landau}\ \emph {et~al.}(1984)\citenamefont {Landau},
  \citenamefont {Bell}, \citenamefont {Kearsley}, \citenamefont {Pitaevskii},
  \citenamefont {Lifshitz},\ and\ \citenamefont {Sykes}}]{Landau:84}%
  \BibitemOpen
  \bibfield  {author} {\bibinfo {author} {\bibfnamefont {Lev~Davidovich}\
  \bibnamefont {Landau}}, \bibinfo {author} {\bibfnamefont {JS}~\bibnamefont
  {Bell}}, \bibinfo {author} {\bibfnamefont {MJ}~\bibnamefont {Kearsley}},
  \bibinfo {author} {\bibfnamefont {LP}~\bibnamefont {Pitaevskii}}, \bibinfo
  {author} {\bibfnamefont {EM}~\bibnamefont {Lifshitz}}, \ and\ \bibinfo
  {author} {\bibfnamefont {JB}~\bibnamefont {Sykes}},\ }\href@noop {} {\emph
  {\bibinfo {title} {Electrodynamics of continuous media}}},\ Vol.~\bibinfo
  {volume} {8}\ (\bibinfo  {publisher} {elsevier},\ \bibinfo {year}
  {1984})\BibitemShut {NoStop}%
\bibitem [{\citenamefont {Cheiwchanchamnangij}\ and\ \citenamefont
  {Lambrecht}(2012)}]{Cheiwchanchamnangij:PRB2012}%
  \BibitemOpen
  \bibfield  {author} {\bibinfo {author} {\bibfnamefont {Tawinan}\ \bibnamefont
  {Cheiwchanchamnangij}}\ and\ \bibinfo {author} {\bibfnamefont {Walter R.~L.}\
  \bibnamefont {Lambrecht}},\ }\bibfield  {title} {\enquote {\bibinfo {title}
  {Quasiparticle band structure calculation of monolayer, bilayer, and bulk
  mos${}_{2}$},}\ }\href {\doibase 10.1103/PhysRevB.85.205302} {\bibfield
  {journal} {\bibinfo  {journal} {Phys. Rev. B}\ }\textbf {\bibinfo {volume}
  {85}},\ \bibinfo {pages} {205302} (\bibinfo {year} {2012})}\BibitemShut
  {NoStop}%
\bibitem [{\citenamefont {Korn}\ \emph {et~al.}(2011)\citenamefont {Korn},
  \citenamefont {Heydrich}, \citenamefont {Hirmer}, \citenamefont
  {Schmutzler},\ and\ \citenamefont {Sch{\"u}ller}}]{Korn:APL11}%
  \BibitemOpen
  \bibfield  {author} {\bibinfo {author} {\bibfnamefont {T.}~\bibnamefont
  {Korn}}, \bibinfo {author} {\bibfnamefont {S.}~\bibnamefont {Heydrich}},
  \bibinfo {author} {\bibfnamefont {M.}~\bibnamefont {Hirmer}}, \bibinfo
  {author} {\bibfnamefont {J.}~\bibnamefont {Schmutzler}}, \ and\ \bibinfo
  {author} {\bibfnamefont {C.}~\bibnamefont {Sch{\"u}ller}},\ }\bibfield
  {title} {\enquote {\bibinfo {title} {Low-temperature photocarrier dynamics in
  monolayer mos2},}\ }\href@noop {} {\bibfield  {journal} {\bibinfo  {journal}
  {Appl. Phys. Lett.}\ }\textbf {\bibinfo {volume} {99}},\ \bibinfo {pages}
  {102109} (\bibinfo {year} {2011})}\BibitemShut {NoStop}%
\bibitem [{\citenamefont {Palummo}\ \emph {et~al.}(2015)\citenamefont
  {Palummo}, \citenamefont {Bernardi},\ and\ \citenamefont
  {Grossman}}]{Palummo:NL15}%
  \BibitemOpen
  \bibfield  {author} {\bibinfo {author} {\bibfnamefont {Maurizia}\
  \bibnamefont {Palummo}}, \bibinfo {author} {\bibfnamefont {Marco}\
  \bibnamefont {Bernardi}}, \ and\ \bibinfo {author} {\bibfnamefont
  {Jeffrey~C.}\ \bibnamefont {Grossman}},\ }\bibfield  {title} {\enquote
  {\bibinfo {title} {Exciton radiative lifetimes in two-dimensional transition
  metal dichalcogenides},}\ }\href@noop {} {\bibfield  {journal} {\bibinfo
  {journal} {Nano Lett.}\ }\textbf {\bibinfo {volume} {15}},\ \bibinfo {pages}
  {2794--2800} (\bibinfo {year} {2015})}\BibitemShut {NoStop}%
\bibitem [{\citenamefont {Amani}\ \emph {et~al.}(2015)\citenamefont {Amani},
  \citenamefont {Lien}, \citenamefont {Kiriya}, \citenamefont {Xiao},
  \citenamefont {Azcatl}, \citenamefont {Noh}, \citenamefont {Madhvapathy},
  \citenamefont {Addou}, \citenamefont {KC}, \citenamefont {Dubey},
  \citenamefont {Cho}, \citenamefont {Wallace}, \citenamefont {Lee},
  \citenamefont {He}, \citenamefont {Ager}, \citenamefont {Zhang},
  \citenamefont {Yablonovitch},\ and\ \citenamefont {Javey}}]{Amani:S15}%
  \BibitemOpen
  \bibfield  {author} {\bibinfo {author} {\bibfnamefont {Matin}\ \bibnamefont
  {Amani}}, \bibinfo {author} {\bibfnamefont {Der-Hsien}\ \bibnamefont {Lien}},
  \bibinfo {author} {\bibfnamefont {Daisuke}\ \bibnamefont {Kiriya}}, \bibinfo
  {author} {\bibfnamefont {Jun}\ \bibnamefont {Xiao}}, \bibinfo {author}
  {\bibfnamefont {Angelica}\ \bibnamefont {Azcatl}}, \bibinfo {author}
  {\bibfnamefont {Jiyoung}\ \bibnamefont {Noh}}, \bibinfo {author}
  {\bibfnamefont {Surabhi~R.}\ \bibnamefont {Madhvapathy}}, \bibinfo {author}
  {\bibfnamefont {Rafik}\ \bibnamefont {Addou}}, \bibinfo {author}
  {\bibfnamefont {Santosh}\ \bibnamefont {KC}}, \bibinfo {author}
  {\bibfnamefont {Madan}\ \bibnamefont {Dubey}}, \bibinfo {author}
  {\bibfnamefont {Kyeongjae}\ \bibnamefont {Cho}}, \bibinfo {author}
  {\bibfnamefont {Robert~M.}\ \bibnamefont {Wallace}}, \bibinfo {author}
  {\bibfnamefont {Si-Chen}\ \bibnamefont {Lee}}, \bibinfo {author}
  {\bibfnamefont {Jr-Hau}\ \bibnamefont {He}}, \bibinfo {author} {\bibfnamefont
  {Joel~W.}\ \bibnamefont {Ager}}, \bibinfo {author} {\bibfnamefont {Xiang}\
  \bibnamefont {Zhang}}, \bibinfo {author} {\bibfnamefont {Eli}\ \bibnamefont
  {Yablonovitch}}, \ and\ \bibinfo {author} {\bibfnamefont {Ali}\ \bibnamefont
  {Javey}},\ }\bibfield  {title} {\enquote {\bibinfo {title} {Near-unity
  photoluminescence quantum yield in mos2},}\ }\href@noop {} {\bibfield
  {journal} {\bibinfo  {journal} {Science}\ }\textbf {\bibinfo {volume}
  {350}},\ \bibinfo {pages} {1065--1068} (\bibinfo {year} {2015})}\BibitemShut
  {NoStop}%
\bibitem [{\citenamefont {Surrente}\ \emph {et~al.}(2016)\citenamefont
  {Surrente}, \citenamefont {Mitioglu}, \citenamefont {Galkowski},
  \citenamefont {Tabis}, \citenamefont {Maude},\ and\ \citenamefont
  {Plochocka}}]{Surrente:PRB16}%
  \BibitemOpen
  \bibfield  {author} {\bibinfo {author} {\bibfnamefont {A.}~\bibnamefont
  {Surrente}}, \bibinfo {author} {\bibfnamefont {A.~A.}\ \bibnamefont
  {Mitioglu}}, \bibinfo {author} {\bibfnamefont {K.}~\bibnamefont {Galkowski}},
  \bibinfo {author} {\bibfnamefont {W.}~\bibnamefont {Tabis}}, \bibinfo
  {author} {\bibfnamefont {D.~K.}\ \bibnamefont {Maude}}, \ and\ \bibinfo
  {author} {\bibfnamefont {P.}~\bibnamefont {Plochocka}},\ }\bibfield  {title}
  {\enquote {\bibinfo {title} {Excitons in atomically thin black phosphorus},}\
  }\href@noop {} {\bibfield  {journal} {\bibinfo  {journal} {Phys. Rev. B}\
  }\textbf {\bibinfo {volume} {93}},\ \bibinfo {pages} {121405} (\bibinfo
  {year} {2016})}\BibitemShut {NoStop}%
\bibitem [{\citenamefont {Suzuura}\ and\ \citenamefont
  {Ando}(2002)}]{Suzuura:PRB02}%
  \BibitemOpen
  \bibfield  {author} {\bibinfo {author} {\bibfnamefont {Hidekatsu}\
  \bibnamefont {Suzuura}}\ and\ \bibinfo {author} {\bibfnamefont {Tsuneya}\
  \bibnamefont {Ando}},\ }\bibfield  {title} {\enquote {\bibinfo {title}
  {Phonons and electron-phonon scattering in carbon nanotubes},}\ }\href
  {\doibase 10.1103/PhysRevB.65.235412} {\bibfield  {journal} {\bibinfo
  {journal} {Phys. Rev. B}\ }\textbf {\bibinfo {volume} {65}},\ \bibinfo
  {pages} {235412} (\bibinfo {year} {2002})}\BibitemShut {NoStop}%
\bibitem [{\citenamefont {Harrison}(1999)}]{Harrison:99}%
  \BibitemOpen
  \bibfield  {author} {\bibinfo {author} {\bibfnamefont {Walter~Ashley}\
  \bibnamefont {Harrison}},\ }\href@noop {} {\emph {\bibinfo {title}
  {Elementary electronic structure}}}\ (\bibinfo  {publisher} {World
  Scientific},\ \bibinfo {year} {1999})\BibitemShut {NoStop}%
\bibitem [{\citenamefont {Wang}\ \emph {et~al.}(2014)\citenamefont {Wang},
  \citenamefont {Kutana},\ and\ \citenamefont {Yakobson}}]{Wang:ADP14}%
  \BibitemOpen
  \bibfield  {author} {\bibinfo {author} {\bibfnamefont {Luqing}\ \bibnamefont
  {Wang}}, \bibinfo {author} {\bibfnamefont {Alex}\ \bibnamefont {Kutana}}, \
  and\ \bibinfo {author} {\bibfnamefont {Boris~I.}\ \bibnamefont {Yakobson}},\
  }\bibfield  {title} {\enquote {\bibinfo {title} {Many-body and spin-orbit
  effects on direct-indirect band gap transition of strained monolayer mos2 and
  ws2},}\ }\href@noop {} {\bibfield  {journal} {\bibinfo  {journal} {Ann.
  Phys.}\ }\textbf {\bibinfo {volume} {526}},\ \bibinfo {pages} {L7--L12}
  (\bibinfo {year} {2014})}\BibitemShut {NoStop}%
\bibitem [{\citenamefont {Appalakondaiah}\ \emph {et~al.}(2012)\citenamefont
  {Appalakondaiah}, \citenamefont {Vaitheeswaran}, \citenamefont {Leb\`egue},
  \citenamefont {Christensen},\ and\ \citenamefont
  {Svane}}]{Appalakondaiah:PRB12}%
  \BibitemOpen
  \bibfield  {author} {\bibinfo {author} {\bibfnamefont {S.}~\bibnamefont
  {Appalakondaiah}}, \bibinfo {author} {\bibfnamefont {G.}~\bibnamefont
  {Vaitheeswaran}}, \bibinfo {author} {\bibfnamefont {S.}~\bibnamefont
  {Leb\`egue}}, \bibinfo {author} {\bibfnamefont {N.~E.}\ \bibnamefont
  {Christensen}}, \ and\ \bibinfo {author} {\bibfnamefont {A.}~\bibnamefont
  {Svane}},\ }\bibfield  {title} {\enquote {\bibinfo {title} {Effect of van der
  waals interactions on the structural and elastic properties of black
  phosphorus},}\ }\href@noop {} {\bibfield  {journal} {\bibinfo  {journal}
  {Phys. Rev. B}\ }\textbf {\bibinfo {volume} {86}},\ \bibinfo {pages} {035105}
  (\bibinfo {year} {2012})}\BibitemShut {NoStop}%
\bibitem [{\citenamefont {Wei}\ and\ \citenamefont {Peng}(2014)}]{Wei:APL14}%
  \BibitemOpen
  \bibfield  {author} {\bibinfo {author} {\bibfnamefont {Qun}\ \bibnamefont
  {Wei}}\ and\ \bibinfo {author} {\bibfnamefont {Xihong}\ \bibnamefont
  {Peng}},\ }\bibfield  {title} {\enquote {\bibinfo {title} {Superior
  mechanical flexibility of phosphorene and few-layer black phosphorus},}\
  }\href@noop {} {\bibfield  {journal} {\bibinfo  {journal} {Appl. Phys.
  Lett.}\ }\textbf {\bibinfo {volume} {104}},\ \bibinfo {pages} {--} (\bibinfo
  {year} {2014})}\BibitemShut {NoStop}%
\bibitem [{\citenamefont {Jiang}\ and\ \citenamefont
  {Park}(2014)}]{Jiang:NC14}%
  \BibitemOpen
  \bibfield  {author} {\bibinfo {author} {\bibfnamefont {Jin-Wu}\ \bibnamefont
  {Jiang}}\ and\ \bibinfo {author} {\bibfnamefont {Harold~S.}\ \bibnamefont
  {Park}},\ }\bibfield  {title} {\enquote {\bibinfo {title} {Negative poisson's
  ratio in single-layer black phosphorus},}\ }\href@noop {} {\bibfield
  {journal} {\bibinfo  {journal} {Nat. Commun.}\ }\textbf {\bibinfo {volume}
  {5}},\ \bibinfo {pages} {4727} (\bibinfo {year} {2014})}\BibitemShut
  {NoStop}%
\bibitem [{\citenamefont {Elahi}\ \emph {et~al.}(2015)\citenamefont {Elahi},
  \citenamefont {Khaliji}, \citenamefont {Tabatabaei}, \citenamefont
  {Pourfath},\ and\ \citenamefont {Asgari}}]{Elahi:PRB15}%
  \BibitemOpen
  \bibfield  {author} {\bibinfo {author} {\bibfnamefont {Mohammad}\
  \bibnamefont {Elahi}}, \bibinfo {author} {\bibfnamefont {Kaveh}\ \bibnamefont
  {Khaliji}}, \bibinfo {author} {\bibfnamefont {Seyed~Mohammad}\ \bibnamefont
  {Tabatabaei}}, \bibinfo {author} {\bibfnamefont {Mahdi}\ \bibnamefont
  {Pourfath}}, \ and\ \bibinfo {author} {\bibfnamefont {Reza}\ \bibnamefont
  {Asgari}},\ }\bibfield  {title} {\enquote {\bibinfo {title} {Modulation of
  electronic and mechanical properties of phosphorene through strain},}\
  }\href@noop {} {\bibfield  {journal} {\bibinfo  {journal} {Phys. Rev. B}\
  }\textbf {\bibinfo {volume} {91}},\ \bibinfo {pages} {115412} (\bibinfo
  {year} {2015})}\BibitemShut {NoStop}%
\bibitem [{\citenamefont {Yue}\ \emph {et~al.}(2012)\citenamefont {Yue},
  \citenamefont {Kang}, \citenamefont {Shao}, \citenamefont {Zhang},
  \citenamefont {Chang}, \citenamefont {Wang}, \citenamefont {Qin},\ and\
  \citenamefont {Li}}]{Yue:PLA12}%
  \BibitemOpen
  \bibfield  {author} {\bibinfo {author} {\bibfnamefont {Qu}~\bibnamefont
  {Yue}}, \bibinfo {author} {\bibfnamefont {Jun}\ \bibnamefont {Kang}},
  \bibinfo {author} {\bibfnamefont {Zhengzheng}\ \bibnamefont {Shao}}, \bibinfo
  {author} {\bibfnamefont {Xueao}\ \bibnamefont {Zhang}}, \bibinfo {author}
  {\bibfnamefont {Shengli}\ \bibnamefont {Chang}}, \bibinfo {author}
  {\bibfnamefont {Guang}\ \bibnamefont {Wang}}, \bibinfo {author}
  {\bibfnamefont {Shiqiao}\ \bibnamefont {Qin}}, \ and\ \bibinfo {author}
  {\bibfnamefont {Jingbo}\ \bibnamefont {Li}},\ }\bibfield  {title} {\enquote
  {\bibinfo {title} {Mechanical and electronic properties of monolayer mos2
  under elastic strain},}\ }\href@noop {} {\bibfield  {journal} {\bibinfo
  {journal} {Phys. Lett. A}\ }\textbf {\bibinfo {volume} {376}},\ \bibinfo
  {pages} {1166 -- 1170} (\bibinfo {year} {2012})}\BibitemShut {NoStop}%
\bibitem [{\citenamefont {{\c C}akir}\ \emph {et~al.}(2014)\citenamefont {{\c
  C}akir}, \citenamefont {Sahin},\ and\ \citenamefont {Peeters}}]{Cakr:PRB14}%
  \BibitemOpen
  \bibfield  {author} {\bibinfo {author} {\bibfnamefont {Deniz}\ \bibnamefont
  {{\c C}akir}}, \bibinfo {author} {\bibfnamefont {Hasan}\ \bibnamefont
  {Sahin}}, \ and\ \bibinfo {author} {\bibfnamefont {Fran{\c c}ois~M.}\
  \bibnamefont {Peeters}},\ }\bibfield  {title} {\enquote {\bibinfo {title}
  {Tuning of the electronic and optical properties of single-layer black
  phosphorus by strain},}\ }\href {\doibase 10.1103/PhysRevB.90.205421}
  {\bibfield  {journal} {\bibinfo  {journal} {Phys. Rev. B}\ }\textbf {\bibinfo
  {volume} {90}},\ \bibinfo {pages} {205421} (\bibinfo {year}
  {2014})}\BibitemShut {NoStop}%
\bibitem [{Note3()}]{Note3}%
  \BibitemOpen
  \bibinfo {note} {For simplicity, we assume the exciton masses $M_{x,y}$ to be
  constants, fixed to their initial values. A more accurate solution of the
  exciton motion using a position-dependent masses shows this is generally a
  rather accurate approximation. We also implicitly assume the strain gradient
  to be effectively adiabatic on the scale of the exciton radius.}\BibitemShut
  {Stop}%
\bibitem [{\citenamefont {Bagaev}\ \emph {et~al.}(1969)\citenamefont {Bagaev},
  \citenamefont {Galkina}, \citenamefont {Gogolin},\ and\ \citenamefont
  {Keldysh}}]{Bagaev:JL69}%
  \BibitemOpen
  \bibfield  {author} {\bibinfo {author} {\bibfnamefont {V.~S.}\ \bibnamefont
  {Bagaev}}, \bibinfo {author} {\bibfnamefont {T.~I.}\ \bibnamefont {Galkina}},
  \bibinfo {author} {\bibfnamefont {O.~V.}\ \bibnamefont {Gogolin}}, \ and\
  \bibinfo {author} {\bibfnamefont {L.~V.}\ \bibnamefont {Keldysh}},\
  }\bibfield  {title} {\enquote {\bibinfo {title} {Motion of electron-hole
  drops in germanium},}\ }\href@noop {} {\bibfield  {journal} {\bibinfo
  {journal} {JETP Lett.}\ }\textbf {\bibinfo {volume} {10}},\ \bibinfo {pages}
  {195} (\bibinfo {year} {1969})}\BibitemShut {NoStop}%
\bibitem [{\citenamefont {Shimizu}(2006)}]{Shimizu:JOL06}%
  \BibitemOpen
  \bibfield  {author} {\bibinfo {author} {\bibfnamefont {Makoto}\ \bibnamefont
  {Shimizu}},\ }\bibfield  {title} {\enquote {\bibinfo {title} {Long-range pair
  transport in graded band gap and its applications},}\ }\href@noop {}
  {\bibfield  {journal} {\bibinfo  {journal} {J. Lumin.}\ }\textbf {\bibinfo
  {volume} {119--120}},\ \bibinfo {pages} {51 -- 54} (\bibinfo {year}
  {2006})}\BibitemShut {NoStop}%
\bibitem [{\citenamefont {Buscema}\ \emph {et~al.}(2015)\citenamefont
  {Buscema}, \citenamefont {Island}, \citenamefont {Groenendijk}, \citenamefont
  {Blanter}, \citenamefont {Steele}, \citenamefont {van~der Zant},\ and\
  \citenamefont {Castellanos-Gomez}}]{Buscema:CSR15}%
  \BibitemOpen
  \bibfield  {author} {\bibinfo {author} {\bibfnamefont {Michele}\ \bibnamefont
  {Buscema}}, \bibinfo {author} {\bibfnamefont {Joshua~O.}\ \bibnamefont
  {Island}}, \bibinfo {author} {\bibfnamefont {Dirk~J.}\ \bibnamefont
  {Groenendijk}}, \bibinfo {author} {\bibfnamefont {Sofya~I.}\ \bibnamefont
  {Blanter}}, \bibinfo {author} {\bibfnamefont {Gary~A.}\ \bibnamefont
  {Steele}}, \bibinfo {author} {\bibfnamefont {Herre S.~J.}\ \bibnamefont
  {van~der Zant}}, \ and\ \bibinfo {author} {\bibfnamefont {Andres}\
  \bibnamefont {Castellanos-Gomez}},\ }\bibfield  {title} {\enquote {\bibinfo
  {title} {Photocurrent generation with two-dimensional van der waals
  semiconductors},}\ }\href {\doibase 10.1039/C5CS00106D} {\bibfield  {journal}
  {\bibinfo  {journal} {Chem. Soc. Rev.}\ }\textbf {\bibinfo {volume} {44}},\
  \bibinfo {pages} {3691--3718} (\bibinfo {year} {2015})}\BibitemShut {NoStop}%
\bibitem [{\citenamefont {Pan}\ \emph {et~al.}(2016)\citenamefont {Pan},
  \citenamefont {Wang}, \citenamefont {Ye}, \citenamefont {Quhe}, \citenamefont
  {Zhong}, \citenamefont {Song}, \citenamefont {Peng}, \citenamefont {Yu},
  \citenamefont {Yang}, \citenamefont {Shi},\ and\ \citenamefont
  {Lu}}]{Pan:COM16}%
  \BibitemOpen
  \bibfield  {author} {\bibinfo {author} {\bibfnamefont {Yuanyuan}\
  \bibnamefont {Pan}}, \bibinfo {author} {\bibfnamefont {Yangyang}\
  \bibnamefont {Wang}}, \bibinfo {author} {\bibfnamefont {Meng}\ \bibnamefont
  {Ye}}, \bibinfo {author} {\bibfnamefont {Ruge}\ \bibnamefont {Quhe}},
  \bibinfo {author} {\bibfnamefont {Hongxia}\ \bibnamefont {Zhong}}, \bibinfo
  {author} {\bibfnamefont {Zhigang}\ \bibnamefont {Song}}, \bibinfo {author}
  {\bibfnamefont {Xiyou}\ \bibnamefont {Peng}}, \bibinfo {author}
  {\bibfnamefont {Dapeng}\ \bibnamefont {Yu}}, \bibinfo {author} {\bibfnamefont
  {Jinbo}\ \bibnamefont {Yang}}, \bibinfo {author} {\bibfnamefont {Junjie}\
  \bibnamefont {Shi}}, \ and\ \bibinfo {author} {\bibfnamefont {Jing}\
  \bibnamefont {Lu}},\ }\bibfield  {title} {\enquote {\bibinfo {title}
  {Monolayer phosphorene--metal contacts},}\ }\href {\doibase
  10.1021/acs.chemmater.5b04899} {\bibfield  {journal} {\bibinfo  {journal}
  {Chemistry of Materials}\ }\textbf {\bibinfo {volume} {28}},\ \bibinfo
  {pages} {2100--2109} (\bibinfo {year} {2016})}\BibitemShut {NoStop}%
\bibitem [{\citenamefont {Dai}\ and\ \citenamefont {Zeng}(2014)}]{Dai:JPCL14}%
  \BibitemOpen
  \bibfield  {author} {\bibinfo {author} {\bibfnamefont {Jun}\ \bibnamefont
  {Dai}}\ and\ \bibinfo {author} {\bibfnamefont {Xiao~Cheng}\ \bibnamefont
  {Zeng}},\ }\bibfield  {title} {\enquote {\bibinfo {title} {Bilayer
  phosphorene: Effect of stacking order on bandgap and its potential
  applications in thin-film solar cells},}\ }\href {\doibase 10.1021/jz500409m}
  {\bibfield  {journal} {\bibinfo  {journal} {J. Phys. Chem. Lett.}\ }\textbf
  {\bibinfo {volume} {5}},\ \bibinfo {pages} {1289--1293} (\bibinfo {year}
  {2014})},\ \bibinfo {note} {pMID: 26274486}\BibitemShut {NoStop}%
\bibitem [{\citenamefont {Deng}\ \emph {et~al.}(2014)\citenamefont {Deng},
  \citenamefont {Luo}, \citenamefont {Conrad}, \citenamefont {Liu},
  \citenamefont {Gong}, \citenamefont {Najmaei}, \citenamefont {Ajayan},
  \citenamefont {Lou}, \citenamefont {Xu},\ and\ \citenamefont
  {Ye}}]{Deng:AN14}%
  \BibitemOpen
  \bibfield  {author} {\bibinfo {author} {\bibfnamefont {Yexin}\ \bibnamefont
  {Deng}}, \bibinfo {author} {\bibfnamefont {Zhe}\ \bibnamefont {Luo}},
  \bibinfo {author} {\bibfnamefont {Nathan~J.}\ \bibnamefont {Conrad}},
  \bibinfo {author} {\bibfnamefont {Han}\ \bibnamefont {Liu}}, \bibinfo
  {author} {\bibfnamefont {Yongji}\ \bibnamefont {Gong}}, \bibinfo {author}
  {\bibfnamefont {Sina}\ \bibnamefont {Najmaei}}, \bibinfo {author}
  {\bibfnamefont {Pulickel~M.}\ \bibnamefont {Ajayan}}, \bibinfo {author}
  {\bibfnamefont {Jun}\ \bibnamefont {Lou}}, \bibinfo {author} {\bibfnamefont
  {Xianfan}\ \bibnamefont {Xu}}, \ and\ \bibinfo {author} {\bibfnamefont
  {Peide~D.}\ \bibnamefont {Ye}},\ }\bibfield  {title} {\enquote {\bibinfo
  {title} {Black phosphorus--monolayer mos2 van der waals heterojunction p--n
  diode},}\ }\href {\doibase 10.1021/nn5027388} {\bibfield  {journal} {\bibinfo
   {journal} {ACS Nano}\ }\textbf {\bibinfo {volume} {8}},\ \bibinfo {pages}
  {8292--8299} (\bibinfo {year} {2014})},\ \bibinfo {note} {pMID:
  25019534}\BibitemShut {NoStop}%
\bibitem [{\citenamefont {Buscema}\ \emph {et~al.}(2014)\citenamefont
  {Buscema}, \citenamefont {Groenendijk}, \citenamefont {Steele}, \citenamefont
  {van~der Zant},\ and\ \citenamefont {Castellanos-Gomez}}]{Buscema:NC14}%
  \BibitemOpen
  \bibfield  {author} {\bibinfo {author} {\bibfnamefont {Michele}\ \bibnamefont
  {Buscema}}, \bibinfo {author} {\bibfnamefont {Dirk~J.}\ \bibnamefont
  {Groenendijk}}, \bibinfo {author} {\bibfnamefont {Gary~A.}\ \bibnamefont
  {Steele}}, \bibinfo {author} {\bibfnamefont {Herre S.~J.}\ \bibnamefont
  {van~der Zant}}, \ and\ \bibinfo {author} {\bibfnamefont {Andres}\
  \bibnamefont {Castellanos-Gomez}},\ }\bibfield  {title} {\enquote {\bibinfo
  {title} {Photovoltaic effect in few-layer black phosphorus pn junctions
  defined by local electrostatic gating},}\ }\href
  {http://dx.doi.org/10.1038/ncomms5651} {\bibfield  {journal} {\bibinfo
  {journal} {Nat Commun}\ }\textbf {\bibinfo {volume} {5}},\ \bibinfo {pages}
  {4651} (\bibinfo {year} {2014})}\BibitemShut {NoStop}%
\bibitem [{\citenamefont {Chaves}\ \emph {et~al.}(2016)\citenamefont {Chaves},
  \citenamefont {Mayers}, \citenamefont {Peeters},\ and\ \citenamefont
  {Reichman}}]{Chaves:PRB16}%
  \BibitemOpen
  \bibfield  {author} {\bibinfo {author} {\bibfnamefont {A.}~\bibnamefont
  {Chaves}}, \bibinfo {author} {\bibfnamefont {M.~Z.}\ \bibnamefont {Mayers}},
  \bibinfo {author} {\bibfnamefont {F.~M.}\ \bibnamefont {Peeters}}, \ and\
  \bibinfo {author} {\bibfnamefont {D.~R.}\ \bibnamefont {Reichman}},\
  }\bibfield  {title} {\enquote {\bibinfo {title} {Theoretical investigation of
  electron-hole complexes in anisotropic two-dimensional materials},}\ }\href
  {\doibase 10.1103/PhysRevB.93.115314} {\bibfield  {journal} {\bibinfo
  {journal} {Phys. Rev. B}\ }\textbf {\bibinfo {volume} {93}},\ \bibinfo
  {pages} {115314} (\bibinfo {year} {2016})}\BibitemShut {NoStop}%
\bibitem [{\citenamefont {Thilagam}(2015)}]{Thilagam:15}%
  \BibitemOpen
  \bibfield  {author} {\bibinfo {author} {\bibfnamefont {A.}~\bibnamefont
  {Thilagam}},\ }\bibfield  {title} {\enquote {\bibinfo {title} {Ultrafast
  exciton relaxation in monolayer transition metal dichalcogenides},}\
  }\href@noop {} {\  (\bibinfo {year} {2015})},\ \Eprint
  {http://arxiv.org/abs/1512.03380} {arXiv:1512.03380} \BibitemShut {NoStop}%
\bibitem [{\citenamefont {Gomes}\ and\ \citenamefont
  {Carvalho}(2015)}]{Gomes:PRB15}%
  \BibitemOpen
  \bibfield  {author} {\bibinfo {author} {\bibfnamefont {L\'{\i}dia~C.}\
  \bibnamefont {Gomes}}\ and\ \bibinfo {author} {\bibfnamefont
  {A.}~\bibnamefont {Carvalho}},\ }\bibfield  {title} {\enquote {\bibinfo
  {title} {Phosphorene analogues: Isoelectronic two-dimensional group-iv
  monochalcogenides with orthorhombic structure},}\ }\href@noop {} {\bibfield
  {journal} {\bibinfo  {journal} {Phys. Rev. B}\ }\textbf {\bibinfo {volume}
  {92}},\ \bibinfo {pages} {085406} (\bibinfo {year} {2015})}\BibitemShut
  {NoStop}%
\bibitem [{\citenamefont {Hu}\ \emph {et~al.}(2015)\citenamefont {Hu},
  \citenamefont {Zhang}, \citenamefont {Sun}, \citenamefont {Xie},
  \citenamefont {Cai},\ and\ \citenamefont {Zeng}}]{Hu:APL15}%
  \BibitemOpen
  \bibfield  {author} {\bibinfo {author} {\bibfnamefont {Yonghong}\
  \bibnamefont {Hu}}, \bibinfo {author} {\bibfnamefont {Shengli}\ \bibnamefont
  {Zhang}}, \bibinfo {author} {\bibfnamefont {Shaofa}\ \bibnamefont {Sun}},
  \bibinfo {author} {\bibfnamefont {Meiqiu}\ \bibnamefont {Xie}}, \bibinfo
  {author} {\bibfnamefont {Bo}~\bibnamefont {Cai}}, \ and\ \bibinfo {author}
  {\bibfnamefont {Haibo}\ \bibnamefont {Zeng}},\ }\bibfield  {title} {\enquote
  {\bibinfo {title} {Gese monolayer semiconductor with tunable direct band gap
  and small carrier effective mass},}\ }\href@noop {} {\bibfield  {journal}
  {\bibinfo  {journal} {App. Phys. Lett.}\ }\textbf {\bibinfo {volume} {107}},\
  \bibinfo {eid} {122107} (\bibinfo {year} {2015})}\BibitemShut {NoStop}%
\bibitem [{\citenamefont {Wu}\ \emph {et~al.}(2014)\citenamefont {Wu},
  \citenamefont {Qian},\ and\ \citenamefont {Li}}]{Wu:NL14}%
  \BibitemOpen
  \bibfield  {author} {\bibinfo {author} {\bibfnamefont {Menghao}\ \bibnamefont
  {Wu}}, \bibinfo {author} {\bibfnamefont {Xiaofeng}\ \bibnamefont {Qian}}, \
  and\ \bibinfo {author} {\bibfnamefont {Ju}~\bibnamefont {Li}},\ }\bibfield
  {title} {\enquote {\bibinfo {title} {Tunable exciton funnel using moir{\'e}
  superlattice in twisted van der waals bilayer},}\ }\href@noop {} {\bibfield
  {journal} {\bibinfo  {journal} {Nano Lett.}\ }\textbf {\bibinfo {volume}
  {14}},\ \bibinfo {pages} {5350--5357} (\bibinfo {year} {2014})}\BibitemShut
  {NoStop}%
\bibitem [{\citenamefont {Alden}\ \emph {et~al.}(2013)\citenamefont {Alden},
  \citenamefont {Tsen}, \citenamefont {Huang}, \citenamefont {Hovden},
  \citenamefont {Brown}, \citenamefont {Park}, \citenamefont {Muller},\ and\
  \citenamefont {McEuen}}]{Alden:PNAS13}%
  \BibitemOpen
  \bibfield  {author} {\bibinfo {author} {\bibfnamefont {Jonathan~S.}\
  \bibnamefont {Alden}}, \bibinfo {author} {\bibfnamefont {Adam~W.}\
  \bibnamefont {Tsen}}, \bibinfo {author} {\bibfnamefont {Pinshane~Y.}\
  \bibnamefont {Huang}}, \bibinfo {author} {\bibfnamefont {Robert}\
  \bibnamefont {Hovden}}, \bibinfo {author} {\bibfnamefont {Lola}\ \bibnamefont
  {Brown}}, \bibinfo {author} {\bibfnamefont {Jiwoong}\ \bibnamefont {Park}},
  \bibinfo {author} {\bibfnamefont {David~A.}\ \bibnamefont {Muller}}, \ and\
  \bibinfo {author} {\bibfnamefont {Paul~L.}\ \bibnamefont {McEuen}},\
  }\bibfield  {title} {\enquote {\bibinfo {title} {Strain solitons and
  topological defects in bilayer graphene},}\ }\href@noop {} {\bibfield
  {journal} {\bibinfo  {journal} {Proc. Nat. Acad. Sci.}\ }\textbf {\bibinfo
  {volume} {110}},\ \bibinfo {pages} {11256--11260} (\bibinfo {year}
  {2013})}\BibitemShut {NoStop}%
\bibitem [{\citenamefont {Yankowitz}\ \emph {et~al.}(2012)\citenamefont
  {Yankowitz}, \citenamefont {Xue}, \citenamefont {Cormode}, \citenamefont
  {Sanchez-Yamagishi}, \citenamefont {Watanabe}, \citenamefont {Taniguchi},
  \citenamefont {Jarillo-Herrero}, \citenamefont {Jacquod},\ and\ \citenamefont
  {LeRoy}}]{Yankowitz:NP12}%
  \BibitemOpen
  \bibfield  {author} {\bibinfo {author} {\bibfnamefont {Matthew}\ \bibnamefont
  {Yankowitz}}, \bibinfo {author} {\bibfnamefont {Jiamin}\ \bibnamefont {Xue}},
  \bibinfo {author} {\bibfnamefont {Daniel}\ \bibnamefont {Cormode}}, \bibinfo
  {author} {\bibfnamefont {Javier~D.}\ \bibnamefont {Sanchez-Yamagishi}},
  \bibinfo {author} {\bibfnamefont {K.}~\bibnamefont {Watanabe}}, \bibinfo
  {author} {\bibfnamefont {T.}~\bibnamefont {Taniguchi}}, \bibinfo {author}
  {\bibfnamefont {Pablo}\ \bibnamefont {Jarillo-Herrero}}, \bibinfo {author}
  {\bibfnamefont {Philippe}\ \bibnamefont {Jacquod}}, \ and\ \bibinfo {author}
  {\bibfnamefont {Brian~J.}\ \bibnamefont {LeRoy}},\ }\bibfield  {title}
  {\enquote {\bibinfo {title} {Emergence of superlattice dirac points in
  graphene on hexagonal boron nitride},}\ }\href@noop {} {\bibfield  {journal}
  {\bibinfo  {journal} {Nat. Phys.}\ }\textbf {\bibinfo {volume} {8}},\
  \bibinfo {pages} {382--386} (\bibinfo {year} {2012})}\BibitemShut {NoStop}%
\bibitem [{\citenamefont {Woods}\ \emph {et~al.}(2014)\citenamefont {Woods},
  \citenamefont {Britnell}, \citenamefont {Eckmann}, \citenamefont {Ma},
  \citenamefont {Lu}, \citenamefont {Guo}, \citenamefont {Lin}, \citenamefont
  {Yu}, \citenamefont {Cao}, \citenamefont {Gorbachev}, \citenamefont
  {Kretinin}, \citenamefont {Park}, \citenamefont {Ponomarenko}, \citenamefont
  {Katsnelson}, \citenamefont {Gornostyrev}, \citenamefont {Watanabe},
  \citenamefont {Taniguchi}, \citenamefont {Casiraghi}, \citenamefont {Gao},
  \citenamefont {Geim},\ and\ \citenamefont {Novoselov}}]{Woods:NP14}%
  \BibitemOpen
  \bibfield  {author} {\bibinfo {author} {\bibfnamefont {C.~R.}\ \bibnamefont
  {Woods}}, \bibinfo {author} {\bibfnamefont {L.}~\bibnamefont {Britnell}},
  \bibinfo {author} {\bibfnamefont {A.}~\bibnamefont {Eckmann}}, \bibinfo
  {author} {\bibfnamefont {R.~S.}\ \bibnamefont {Ma}}, \bibinfo {author}
  {\bibfnamefont {J.~C.}\ \bibnamefont {Lu}}, \bibinfo {author} {\bibfnamefont
  {H.~M.}\ \bibnamefont {Guo}}, \bibinfo {author} {\bibfnamefont
  {X.}~\bibnamefont {Lin}}, \bibinfo {author} {\bibfnamefont {G.~L.}\
  \bibnamefont {Yu}}, \bibinfo {author} {\bibfnamefont {Y.}~\bibnamefont
  {Cao}}, \bibinfo {author} {\bibfnamefont {R.~V.}\ \bibnamefont {Gorbachev}},
  \bibinfo {author} {\bibfnamefont {A.~V.}\ \bibnamefont {Kretinin}}, \bibinfo
  {author} {\bibfnamefont {J.}~\bibnamefont {Park}}, \bibinfo {author}
  {\bibfnamefont {L.~A.}\ \bibnamefont {Ponomarenko}}, \bibinfo {author}
  {\bibfnamefont {M.~I.}\ \bibnamefont {Katsnelson}}, \bibinfo {author}
  {\bibfnamefont {Yu.~N.}\ \bibnamefont {Gornostyrev}}, \bibinfo {author}
  {\bibfnamefont {K.}~\bibnamefont {Watanabe}}, \bibinfo {author}
  {\bibfnamefont {T.}~\bibnamefont {Taniguchi}}, \bibinfo {author}
  {\bibfnamefont {C.}~\bibnamefont {Casiraghi}}, \bibinfo {author}
  {\bibfnamefont {H-J.}\ \bibnamefont {Gao}}, \bibinfo {author} {\bibfnamefont
  {A.~K.}\ \bibnamefont {Geim}}, \ and\ \bibinfo {author} {\bibfnamefont
  {K.~S.}\ \bibnamefont {Novoselov}},\ }\bibfield  {title} {\enquote {\bibinfo
  {title} {Commensurate-incommensurate transition in graphene on hexagonal
  boron nitride},}\ }\href@noop {} {\bibfield  {journal} {\bibinfo  {journal}
  {Nat. Phys.}\ }\textbf {\bibinfo {volume} {10}},\ \bibinfo {pages} {451--456}
  (\bibinfo {year} {2014})}\BibitemShut {NoStop}%
\bibitem [{\citenamefont {San-Jose}\ \emph {et~al.}(2014)\citenamefont
  {San-Jose}, \citenamefont {Guti\'errez-Rubio}, \citenamefont {Sturla},\ and\
  \citenamefont {Guinea}}]{San-Jose:PRB14}%
  \BibitemOpen
  \bibfield  {author} {\bibinfo {author} {\bibfnamefont {Pablo}\ \bibnamefont
  {San-Jose}}, \bibinfo {author} {\bibfnamefont {A.}~\bibnamefont
  {Guti\'errez-Rubio}}, \bibinfo {author} {\bibfnamefont {Mauricio}\
  \bibnamefont {Sturla}}, \ and\ \bibinfo {author} {\bibfnamefont {Francisco}\
  \bibnamefont {Guinea}},\ }\bibfield  {title} {\enquote {\bibinfo {title}
  {Spontaneous strains and gap in graphene on boron nitride},}\ }\href@noop {}
  {\bibfield  {journal} {\bibinfo  {journal} {Phys. Rev. B}\ }\textbf {\bibinfo
  {volume} {90}},\ \bibinfo {pages} {075428} (\bibinfo {year}
  {2014})}\BibitemShut {NoStop}%
\bibitem [{\citenamefont {Yankowitz}\ \emph {et~al.}(2016)\citenamefont
  {Yankowitz}, \citenamefont {Watanabe}, \citenamefont {Taniguchi},
  \citenamefont {San-Jose},\ and\ \citenamefont {LeRoy}}]{Yankowitz:16}%
  \BibitemOpen
  \bibfield  {author} {\bibinfo {author} {\bibfnamefont {Matthew}\ \bibnamefont
  {Yankowitz}}, \bibinfo {author} {\bibfnamefont {K.}~\bibnamefont {Watanabe}},
  \bibinfo {author} {\bibfnamefont {T.}~\bibnamefont {Taniguchi}}, \bibinfo
  {author} {\bibfnamefont {Pablo}\ \bibnamefont {San-Jose}}, \ and\ \bibinfo
  {author} {\bibfnamefont {Brian~J.}\ \bibnamefont {LeRoy}},\ }\bibfield
  {title} {\enquote {\bibinfo {title} {Pressure-induced commensurate stacking
  of graphene on boron nitride},}\ }\href@noop {} {\  (\bibinfo {year}
  {2016})},\ \Eprint {http://arxiv.org/abs/1603.03244} {arXiv:1603.03244}
  \BibitemShut {NoStop}%
\bibitem [{\citenamefont {Tran}\ \emph {et~al.}(2015)\citenamefont {Tran},
  \citenamefont {Fei},\ and\ \citenamefont {Yang}}]{Tran:2M15}%
  \BibitemOpen
  \bibfield  {author} {\bibinfo {author} {\bibfnamefont {Vy}~\bibnamefont
  {Tran}}, \bibinfo {author} {\bibfnamefont {Ruixiang}\ \bibnamefont {Fei}}, \
  and\ \bibinfo {author} {\bibfnamefont {Li}~\bibnamefont {Yang}},\ }\bibfield
  {title} {\enquote {\bibinfo {title} {Quasiparticle energies, excitons, and
  optical spectra of few-layer black phosphorus},}\ }\href@noop {} {\bibfield
  {journal} {\bibinfo  {journal} {2D Materials}\ }\textbf {\bibinfo {volume}
  {2}},\ \bibinfo {pages} {044014} (\bibinfo {year} {2015})}\BibitemShut
  {NoStop}%
\bibitem [{\citenamefont {Zaslow}\ and\ \citenamefont
  {Zandler}(1967)}]{Zaslow:AJOP67}%
  \BibitemOpen
  \bibfield  {author} {\bibinfo {author} {\bibfnamefont {B}~\bibnamefont
  {Zaslow}}\ and\ \bibinfo {author} {\bibfnamefont {Melvin~E}\ \bibnamefont
  {Zandler}},\ }\bibfield  {title} {\enquote {\bibinfo {title} {Two-dimensional
  analog to the hydrogen atom},}\ }\href@noop {} {\bibfield  {journal}
  {\bibinfo  {journal} {American Journal of Physics}\ }\textbf {\bibinfo
  {volume} {35}},\ \bibinfo {pages} {1118--1119} (\bibinfo {year}
  {1967})}\BibitemShut {NoStop}%
\bibitem [{\citenamefont {Gabor}\ \emph {et~al.}(2011)\citenamefont {Gabor},
  \citenamefont {Song}, \citenamefont {Ma}, \citenamefont {Nair}, \citenamefont
  {Taychatanapat}, \citenamefont {Watanabe}, \citenamefont {Taniguchi},
  \citenamefont {Levitov},\ and\ \citenamefont {Jarillo-Herrero}}]{Gabor648}%
  \BibitemOpen
  \bibfield  {author} {\bibinfo {author} {\bibfnamefont {Nathaniel~M.}\
  \bibnamefont {Gabor}}, \bibinfo {author} {\bibfnamefont {Justin C.~W.}\
  \bibnamefont {Song}}, \bibinfo {author} {\bibfnamefont {Qiong}\ \bibnamefont
  {Ma}}, \bibinfo {author} {\bibfnamefont {Nityan~L.}\ \bibnamefont {Nair}},
  \bibinfo {author} {\bibfnamefont {Thiti}\ \bibnamefont {Taychatanapat}},
  \bibinfo {author} {\bibfnamefont {Kenji}\ \bibnamefont {Watanabe}}, \bibinfo
  {author} {\bibfnamefont {Takashi}\ \bibnamefont {Taniguchi}}, \bibinfo
  {author} {\bibfnamefont {Leonid~S.}\ \bibnamefont {Levitov}}, \ and\ \bibinfo
  {author} {\bibfnamefont {Pablo}\ \bibnamefont {Jarillo-Herrero}},\ }\bibfield
   {title} {\enquote {\bibinfo {title} {Hot carrier{\textendash}assisted
  intrinsic photoresponse in graphene},}\ }\href {\doibase
  10.1126/science.1211384} {\bibfield  {journal} {\bibinfo  {journal}
  {Science}\ }\textbf {\bibinfo {volume} {334}},\ \bibinfo {pages} {648--652}
  (\bibinfo {year} {2011})}\BibitemShut {NoStop}%
\bibitem [{\citenamefont {Perdew}(1985)}]{Perdew:IJQC85}%
  \BibitemOpen
  \bibfield  {author} {\bibinfo {author} {\bibfnamefont {John~P.}\ \bibnamefont
  {Perdew}},\ }\bibfield  {title} {\enquote {\bibinfo {title} {Density
  functional theory and the band gap problem},}\ }\href@noop {} {\bibfield
  {journal} {\bibinfo  {journal} {Int. J. Quant. Chem.}\ }\textbf {\bibinfo
  {volume} {28}},\ \bibinfo {pages} {497--523} (\bibinfo {year}
  {1985})}\BibitemShut {NoStop}%
\bibitem [{\citenamefont {Qiu}\ \emph {et~al.}(2013)\citenamefont {Qiu},
  \citenamefont {da~Jornada},\ and\ \citenamefont {Louie}}]{Qiu:PRL13}%
  \BibitemOpen
  \bibfield  {author} {\bibinfo {author} {\bibfnamefont {Diana~Y.}\
  \bibnamefont {Qiu}}, \bibinfo {author} {\bibfnamefont {Felipe~H.}\
  \bibnamefont {da~Jornada}}, \ and\ \bibinfo {author} {\bibfnamefont
  {Steven~G.}\ \bibnamefont {Louie}},\ }\bibfield  {title} {\enquote {\bibinfo
  {title} {Optical spectrum of ${\mathrm{mos}}_{2}$: Many-body effects and
  diversity of exciton states},}\ }\href@noop {} {\bibfield  {journal}
  {\bibinfo  {journal} {Phys. Rev. Lett.}\ }\textbf {\bibinfo {volume} {111}},\
  \bibinfo {pages} {216805} (\bibinfo {year} {2013})}\BibitemShut {NoStop}%
\end{thebibliography}%

\end{document}